\documentclass[aps, pra, twocolumn, superscriptaddress, showpacs, tightenlines]{revtex4}
\usepackage{graphicx}
\usepackage{amssymb}
\usepackage{amsmath}
\usepackage{amsfonts}
\usepackage{latexsym,bm}
\usepackage{placeins}

\begin{document}


\title{Negative Refraction in M\"{o}bius Molecules}
\author{Y. N. Fang}
\affiliation{Beijing Computational Science Research Center, Beijing 100084, China}
\affiliation{State Key Laboratory of Theoretical Physics, Institute of Theoretical
Physics, Chinese Academy of Sciences, and University of the Chinese
Academy of Sciences, Beijing 100190, China}
\affiliation{Synergetic Innovation Center of Quantum Information and Quantum
Physics, University of Science and Technology of China, Hefei, Anhui
230026, China}

\author{Yao Shen}
\affiliation{School of Criminal Science and Technology, People's Public
Security University of China, Beijing 100038, China}

\author{Qing Ai}
\email{aiqing@bnu.edu.cn}
\affiliation{Department of Physics, Beijing Normal University, Beijing
100875, China}

\author{C. P. Sun}
\email{cpsun@csrc.ac.cn}
\affiliation{Beijing Computational Science Research Center, Beijing 100084, China}
\affiliation{Synergetic Innovation Center of Quantum Information and Quantum
Physics, University of Science and Technology of China, Hefei, Anhui
230026, China}

\begin{abstract}
We theoretically show the negative refraction existing in M\"{o}bius
molecules. The negative refractive index is induced by the non-trivial
topology of the molecules. With the M\"{o}bius boundary condition,
the effective electromagnetic fields felt by the electron in a M\"{o}bius
ring is spatially inhomogeneous. In this regard, the $D_{N}$ symmetry
is broken in M\"{o}bius molecules and thus the magnetic response
is induced through the effective magnetic field. Our findings open
up a new architecture for negative refractive index materials based
on the non-trivial topology of M\"{o}bius molecules.
\end{abstract}


\maketitle


\section{Introduction}

\label{Introduction}

It is recognized that materials with simultaneously negative permittivity
and permeability also support the coherent propagation of electromagnetic
fields \cite{Veselago68}, like the usual case where both permittivity
and permeability are positive. Due to a reversed phase velocity with
respect to the group velocity \cite{Pendry04}, the boundary condition
makes the light to bend in an unexpected direction as if the refractive
index appears negative according to the Snell's law. Hence, the effect
is termed as negative refraction.

Negative refractive index materials are very promising, e.g., in achieving
the electromagnetic field cloaking \cite{Leonhardt06,Pendry06}, facilitating
the sub-wavelength imaging \cite{Pendry00}, and crime scene investigation
\cite{Shen16}. However, such materials do not naturally exist for
the absence of magnetic response at the same frequency as electric
response. To overcome this difficulty, delicately designed metamaterials
have been proposed and tested. Metamaterials rely on structural designing
of the sub-wavelength unit cells to tune the electromagnetic resonant
characteristic \cite{Pendry06}. A key element commonly involved in
such artificial architectures is the configuration of split-ring resonator
\cite{Pendry99,Smith00,Padilla06,Chang10}. This structure is analogous
to an LC resonator with characteristic frequency $\omega_{0}\sim1/\sqrt{LC}$
with $L$ and $C$ being self-inductance and capacitance of the resonator,
which is tunable by engineering. A metamaterial with split-ring resonators
can negatively respond to the magnetic field when $\omega\gtrsim\omega_{0}$
\cite{Padilla06}. Although the negative refraction has been demonstrated
at the visible wavelength, manufacturing of microstructure at the
size of $30$-$100$nm is required \cite{McPhedran11}. Such technically
requesting fabrication makes it still an open challenge to scale the
materials to 3D bulk.

Instead of fabricating those delicately designed structures from the
conventional ``top-down'' approach, an attractive alternative is
to generate the dielectric media with self-assembled functional
atoms or molecules \cite{Chen05,Thommen06,Shen14}. This ``bottom-up''
approach is particularly interesting due to the small size of molecules
and their associated quantum effects. For example, by utilizing the
quantum interference among multi-level atoms, it is possible to suppress
the absorption meanwhile keep reasonable optical response \cite{Orth13}.
We notice that the molecular ring can not effectively respond to
the magnetic field, because there does not exist any split to make
the symmetry breaking as shown in Refs.~\cite{Brechtefeld06,Shen14} and Appendix~\ref{Appendix1}.

In this paper, instead of relying on the split-ring configuration,
we describe a potential negative-refracting dielectric medium with
the M\"{o}bius molecules \cite{Heilbronner64,Zhao09}. Such molecules
were theoretically proposed \cite{Heilbronner64} and had been fabricated
\cite{Ajami03,Yoneda14,Balzani08} in several experiments. Compared
with other molecular-ring-like annulenes, the symmetry of M\"{o}bius
molecule is lowered by its boundary condition, i.e., the usual $D_{N}$
symmetry is broken here to $C_{2}$ \cite{Zhao09}. Similar to a superconducting
ring with magnetic flux going through \cite{Byers61}, this particular
boundary condition for M\"{o}bius molecule could be canceled by some
local unitary transformations. After the transformation, the motion
of an electron in the M\"{o}bius ring is subject to an effective
spatially-inhomogeneous magnetic field, even though the original system
is exposed to a physically uniform electromagnetic field \cite{Zhao09}.
Both electric and magnetic responses can be induced at the same transitions
and thus both negative permittivity and permeability can be observed
at the same frequencies. In this sense, the negative refraction in
M\"{o}bius molecules is intrinsically caused by the non-trivial topology
of the molecules. Motivated by this discovery, we propose realizing
a new kind of metamaterials based on the M\"{o}bius molecules.

This paper is organized as follows. In the subsequent section, the
M\"{o}bius molecule is briefly introduced following Ref.~\cite{Zhao09}.
In Sec.~\ref{TSR}, the allowed transitions between different eigenstates
of a M\"{o}bius molecule induced by applied electromagnetic fields
are discussed under dipole approximation. Then, in Sec.~\ref{Perm}
we investigate the possibility of having simultaneously negative
permittivity and permeability in a medium consisting of non-interacting
M\"{o}bius molecules. In Sec.~\ref{sec:NegRefMedInterface}, the
negative refraction is analyzed at a planar medium interface for two
different incident conditions. And discussions regarding loss and
bandwidth of negative refraction are also given. Finally, the main
results are summarized in Sec.~VI.

\begin{figure}
\begin{centering}
\includegraphics[bb=0bp 0bp 1128bp 519bp,width=9.5cm]{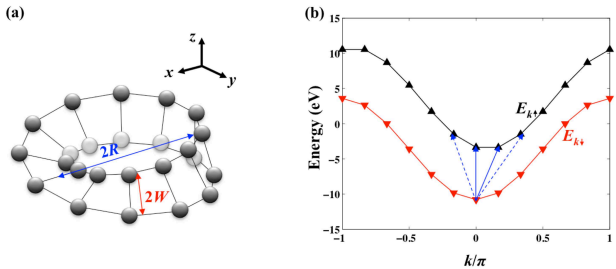}
\par\end{centering}

\caption{(color online) (a) Schematic of a molecular ring with M\"{o}bius
topology: carbon atoms are shown by black circles, with the bonding
atoms linked together. (b) Energy spectrum $E_{k\sigma}$ of the molecular
ring. Two energy bands are denoted by their different pseudo spin
labels $\sigma=\uparrow$ and $\downarrow$. Allowed inter-band transitions
from the ground state are indicated by the colored arrows, where dashed
($x$ and $y$) and solid ($x$, $y$, and $z$) arrows indicate the
possible field polarizations. \label{fig:Schematic}}
\end{figure}

\section{The M\"{o}bius Molecular Ring}

\label{MolecularRing}

The M\"{o}bius molecular ring has several physical realizations,
e.g., through graphene \cite{Guo09} and the non-conjugated molecules
\cite{Walba93}. For a general double-ring system with $2N$ atoms,
as shown in Fig.~\ref{fig:Schematic}(a), the H\"{u}ckel Hamiltonian
for a single electron reads \cite{Salem72,Zhao09}
\begin{equation}
H=\sum_{j=0}^{N-1}\left[\mathbf{A}_{j}^{\dagger}\mathbf{M}\mathbf{A}_{j}-\xi\left(\mathbf{A}_{j}^{\dagger}\mathbf{A}_{j+1}+\mathrm{h.c.}\right)\right],
\label{Hamiltonian}
\end{equation}
where
\begin{eqnarray}
\mathbf{A}_{j} & = & \left[\begin{array}{c}
a_{j}\\
b_{j}
\end{array}\right],\\
\mathbf{M} & = & \left[\begin{array}{cc}
\epsilon & -V\\
-V & -\epsilon
\end{array}\right].
\end{eqnarray}
Here, the fermionic operators $a_{j}^{\dagger}$($b_{j}^{\dagger}$)
create a localized atomic-orbital $|\phi_{j+}\rangle$ ($|\phi_{j-}\rangle$)
at the $j$th nuclear site of the A (B) ring respectively; $2\epsilon$
describes the on-site energy difference between atoms of two rings;
The inter (intra) -ring resonance integral is denoted by $V$ ($\xi$).
Hereafter, we consider a special case where all atoms are of the same
species, i.e., $\epsilon=0$ \cite{Salem72}, as indicated in Fig.~\ref{fig:Schematic}(a).

The difference between the M\"{o}bius molecular ring and the common
chemical annulenes lies in the boundary condition \cite{Zhao09}.
Here, the $N$th nucleus in the A ring is exactly the $0$th nucleus
of the B ring. Therefore,
\begin{eqnarray}
a_{0} & = & b_{N},\\
b_{0} & = & a_{N}
\end{eqnarray}
imply that the operators do not obey the periodical boundary condition.
Further calculations in Appendix~\ref{Appendix1} demonstrate that the vanishing magnetic dipole
for the perfect ring results in the absence of negative refraction.

With a local unitary transformation $\mathbf{U}_{j}$ (see Ref.~\cite{Zhao09} and Appendix~\ref{Appendix2}),
the Hamiltonian becomes
\begin{equation}
H=\sum_{j=0}^{N-1}[\mathbf{B}_{j}^{\dagger}V\mathbf{\sigma}_{z}\mathbf{B}_{j}-\xi(\mathbf{B}_{j}^{\dagger}\mathbf{Q}\mathbf{B}_{j+1}+\mathrm{h.c.})],\label{eq:H}
\end{equation}
where
\begin{eqnarray}
\mathbf{B}_{j} & = & \mathbf{U}_{j}\mathbf{A}_{j}\equiv\left[\begin{array}{c}
c_{j\uparrow}\\
c_{j\downarrow}
\end{array}\right],\\
\mathbf{Q} & = & \left[\begin{array}{cc}
e^{i\delta/2} & 0\\
0 & 1
\end{array}\right],\\
\delta & = & 2\pi/N,
\end{eqnarray}
the Pauli matrix $\sigma_{z}$ is written in the pseudo spin space,
i.e., $\mathbf{B}_{j}^{\dagger}\sigma_{z}\mathbf{B}_{j}=c_{j\uparrow}^{\dagger}c_{j\uparrow}-c_{j\downarrow}^{\dagger}c_{j\downarrow}$.
We emphasize that the new operators $c_{j\sigma}$'s satisfy the periodical
boundary condition. Equation~(\ref{eq:H}) represents a pseudo spin
in a fictitious ring \cite{Zhao09}. For the spin-up state, the ring
is further penetrated by a perpendicular homogeneous magnetic field,
as indicated by the phase factor $\exp\left(i\delta/2\right)$ in
the hopping term. More insight on the non-trivial boundary condition
stems from the corresponding energy spectrum, cf. Fig.~\ref{fig:Schematic}(b) and Appendix~\ref{Appendix2}:
\begin{eqnarray}
E_{k\uparrow} & = & V-2\xi\cos(k-\frac{\delta}{2}),\\
E_{k\downarrow} & = & -V-2\xi\cos k,
\end{eqnarray}
which are obtained by performing the Fourier transform on $\mathbf{B}_{j}$
with the resulting transformed operator denoted as $\mathbf{C}_{k}$
and $k=l\delta$ ($l=0,1,\cdots N-1$). Compared with the spectrum
of a topologically trivial annulene, the effective
magnetic flux results in a shift of the $E_{k\uparrow}$ band such
that the two lowest states become degenerate. Similar shifting behavior
has also been suggested for a cyclic polyene under a real magnetic
field \cite{Salem72}.

\section{Transition selection rules}

\label{TSR}

The general interaction Hamiltonian of a molecule with external electromagnetic
field is complicated. But due to the small size of a molecule (typically
around a few nm) with respect to wavelength the dipole approximation
is applicable. In the linear response regime, non-vanishing electromagnetic
response implies that the corresponding dipole operators have off-diagonal
elements between different eigenstates of $H$. Hence it is useful
to analyze the transition selection rules under the dipole approximation.
We investigate the dipole transition selection rule of the M\"{o}bius
molecular ring under the perturbation of an external oscillating electric
field with amplitude $\vec{E}_{0}$ and frequency $\omega$. The interaction
Hamiltonian is written under the dipole approximation as
\begin{equation}
H_{E}^{\prime}=-\vec{d}\cdot\vec{E}_{0}\cos\omega t,\label{eq:He}
\end{equation}
where the electric dipole operator $\vec{d}=-e\vec{r}$ and $\vec{r}$
is the position operator for the electron.

The transition selection rule can be inferred from the matrix elements
of $H_{E}^{\prime}$ between the eigenstates of $H$, which are the
linear combinations of the atomic orbitals. In the following
we ignore the overlap integrals from different atomic orbitals and
assume \cite{Salem72,Shen14}
\begin{equation}
\left\langle \phi_{js}\left|\vec{r}\right|\phi_{j^{\prime}s^{\prime}}\right\rangle =\delta_{jj^{\prime}}\delta_{ss^{\prime}}\vec{R}_{js},
\end{equation}
where $\vec{R}_{j+}$($\vec{R}_{j-}$) denotes the position vector of the $j$th
nuclear site at the A(B) ring. For the M\"{o}bius molecular ring
with radius $R$ and width $4W$, the nuclear positions are explicitly
written in the molecular coordinate system, cf. Fig.~\ref{fig:Schematic}(a)
and Ref.~\cite{Zhao09}, as
\begin{eqnarray}
\vec{R}_{j\pm} & \!\!=\!\! & (R\pm W\sin\frac{\varphi_{j}}{2})\cos\varphi_{j}\hat{e}_{x}+(R\pm W\sin\frac{\varphi_{j}}{2})\sin\varphi_{j}\hat{e}_{y}\nonumber \\
 &  & \pm W\cos\frac{\varphi_{j}}{2}\hat{e}_{z},
\end{eqnarray}
where
\begin{equation}
\varphi_{j}=j\delta \label{phi}
\end{equation}
is a rotating angle which locates atoms on the rings.

The transition selection rules for the electric dipole operator are
briefly summarized as, cf. Appendix~\ref{Appendix3},
\begin{equation}
\begin{array}{c}
\left|k,\downarrow\right\rangle \overset{x,y,z}{\rightleftharpoons}\left|k,\uparrow\right\rangle ,\left|k,\downarrow\right\rangle \overset{x,y}{\rightleftharpoons}\left|k+2\delta,\uparrow\right\rangle ,\\
\left|k,\sigma\right\rangle \overset{x,y}{\rightleftharpoons}\left|k\pm\delta,\sigma'\right\rangle ,\left|k,\downarrow\right\rangle \overset{z}{\rightleftharpoons}\left|k+\delta,\uparrow\right\rangle .
\end{array}\label{eq:Dk0}
\end{equation}
Here, $\left|k,\sigma\right\rangle $ denotes an eigenstate of $H$
with energy $E_{k\sigma}$. The superscripts $x,y,z$ indicate the
electric field polarizations. Those allowed inter-band transitions
are schematically shown in Fig.~\ref{fig:Schematic}(b).

Since the negative refraction depends on the simultaneously negative
permittivity and permeability, we further investigate the transition
selection rules for the magnetic dipole $\vec{m}$. For molecular
systems the current is restricted to the chemical bond, thus $\vec{m}$
is written as \cite{Ceulemans98} $\vec{m}=\sum_{\left\langle i,j\right\rangle }J_{ij}\vec{S}_{ij}$,
where $\vec{S}_{ij}$ is an effective area specified below. The bond
current $J_{ij}$ flows from the $i$th atom to the $j$th atom, and
by the tight-binding approximation the summation only runs over bonded-atoms
pairs. The operator corresponding to the bond current is given by
$\hat{J}_{ij}=ie\beta_{ij}a_{i}^{\dagger}a_{j}+\mathrm{h.c.}$ \cite{Ceulemans98},
where $\beta_{ij}$ is the resonance integral. For the M\"{o}bius
molecular ring, $\beta_{ij}$ is either $\xi$ or $V$, depending
on whether the atoms are on the same ring or not. Consequently, the
magnetic dipole operator is explicitly given as
\begin{equation}
\vec{m}=-e\sum_{j}[\mathbf{A}_{j}^{\dagger}(V\vec{S}_{j,j}^{+-}\sigma_{y})\mathbf{A}_{j}+\xi(i\mathbf{A}_{j}^{\dagger}\vec{\mathbf{N}}_{j}\mathbf{A}_{j+1}+\mathrm{h.c.})]
\end{equation}
with the effective area
\begin{equation}
\vec{S}_{ij}^{\alpha\beta}\equiv\frac{1}{2}\vec{R}_{i\alpha}\times\vec{R}_{j\beta}
\end{equation}
($\alpha,\beta=\pm$) and the vectorial matrix
\begin{equation}
\vec{\mathbf{N}}_{j}=\left[\begin{array}{cc}
\vec{S}_{j,j+1}^{++} & 0\\
0 & \vec{S}_{j,j+1}^{--}
\end{array}\right].
\end{equation}

Under the dipole approximation, the interaction Hamiltonian describing
the coupling of the M\"{o}bius molecular ring to an external oscillating
magnetic field with amplitude $\vec{B}_{0}$ and frequency $\omega$
is similar to Eq.~(\ref{eq:He}):
\begin{equation}
H{}_{B}^{\prime}=-\vec{m}\cdot\vec{B}_{0}\cos\omega t. \label{eq:Hb}
\end{equation}
The explicit expressions of the matrix elements are complicated in
the case of magnetic perturbation and thus we list them in Appendix~\ref{Appendix3}. Through straightforward calculations, we find the following transition selection rules as illustrated in Fig.~\ref{fig:Schematic}(b):
\begin{equation}
\begin{array}{c}
\left|k,\downarrow\right\rangle \!\overset{xyz}{\rightleftharpoons}\!\left|k,\uparrow\right\rangle ,\mbox{ }\left|k,\downarrow\right\rangle \!\overset{xy}{\rightleftharpoons}\!\left|k+2\delta,\uparrow\right\rangle ,\\
\left|k,\uparrow\right\rangle \!\overset{xy}{\rightleftharpoons}\!\left|k+\delta,\downarrow\right\rangle ,\left|k,\downarrow\right\rangle \!\overset{xyz}{\rightleftharpoons}\!\left|k+\delta,\uparrow\right\rangle ,
\end{array}\label{eq:Mk0}
\end{equation}
where only the inter-band transitions are shown since in the limit
of large $N$, the spectra $E_{k\sigma}$'s become continuous with
respect to $k$ and thus the inter-band transitions are more relevant
to the response at high frequency as compared to the intra-band transitions.
A comparison between Eq.~(\ref{eq:Mk0}) and Eq.~(\ref{eq:Dk0})
indicates that the selection rules for the inter-band transitions
are almost the same for both electric and magnetic couplings.

\section{Permittivity and permeability}

\label{Perm}

In order to realize the negative refraction, the simultaneously negative
permittivity $\overleftrightarrow{\varepsilon_{r}}$ and permeability
$\overleftrightarrow{\mu_{r}}$ are required \cite{Veselago68}. To
evaluate those quantities, let us consider a negative index medium
realized by single crystal of M\"{o}bius molecules, with the same
orientation for each molecule. Although the permittivity
and permeability are originally treated as scalars, for anisotropic
medium they could generally be second order tensors that depend on
the frequency $\omega$ of external fields $\vec{E}$ and $\vec{H}$.
$\overleftrightarrow{\varepsilon_{r}}$ and $\overleftrightarrow{\mu_{r}}$
are calculated by considering that the electric displacement $\vec{D}$
and magnetic induction $\vec{B}$ are related to the polarization
$\vec{P}$ and the magnetization $\vec{M}$ of the medium \cite{Jackson99,Shen14}:
\begin{eqnarray}
\vec{D} & = & \varepsilon_{0}\overleftrightarrow{\varepsilon_{r}}\vec{E}=\varepsilon_{0}\vec{E}+\vec{P},\\
\vec{B} & = & \mu_{0}\overleftrightarrow{\mu_{r}}\vec{H}=\mu_{0}\vec{H}+\mu_{0}\vec{M}.
\end{eqnarray}
Here, $\vec{P}$ and $\vec{M}$ are quantum mechanical averaging of
the electric and magnetic dipole operators in the \textit{perturbed}
molecular ground state. In the linear response regime, applying the
Green-Kubo formula yields \cite{Kubo85}, cf. Appendix~\ref{Appendix4},
\begin{eqnarray}
\vec{P} & = & -\frac{1}{\hbar\upsilon_{0}}\mathrm{Re}\sum_{k,\sigma}\mbox{}'\frac{\vec{d}_{g,k\sigma}(\vec{d}_{k\sigma,g}\cdot\vec{E})}{\omega-\Delta_{k\sigma}+i\gamma},\label{eq:P}\\
\vec{M} & = & -\frac{\mu_{0}}{\hbar\upsilon_{0}}\mathrm{Re}\sum_{k,\sigma}\mbox{}'\frac{\vec{m}_{g,k\sigma}(\vec{m}_{k\sigma,g}\cdot\vec{H})}{\omega-\Delta_{k\sigma}+i\gamma}.\label{eq:M}
\end{eqnarray}
Here, $\Delta_{k\sigma}=E_{k\sigma}-E_{g}$ is the molecular resonant
transition frequency between the excited and ground states of Hamiltonian
$H$. $O_{g,k\sigma}=\langle g|O|k,\sigma\rangle$ is the matrix element
of operator $O$ between the unperturbed molecular ground state $|g\rangle$
and excited state $|k,\sigma\rangle$. The ground state is excluded
in the summation, as indicated by the prime. The lifetimes $\tau=1/\gamma$
of the excited states are assumed to be identical. $\upsilon_{0}$
is the volume occupied by a M\"{o}bius molecule in the medium. In
the above calculation, the inter-molecule interactions have been neglected,
i.e., the polarization and magnetization are assumed to be the contribution
by a collection of non-interacting particles. This approximation is
valid either in the low density limit or for medium with crystal
structure, e.g., Zinc-Blende \cite{Jackson99}. Also, we have verified
that both the central frequency and bandwidth of negative refraction are not substantially
modified when the inter-molecular interaction is considered at the
level of Lorentz local field theory~\cite{Marques08} as proven in Appendix \ref{Appendix_Add}.

Because $\overleftrightarrow{\varepsilon_{r}}\simeq1-\mathrm{O}(|\vec{P}|/|\vec{E}|)$
and $\overleftrightarrow{\mu_{r}}\simeq1-\mathrm{O}(|\vec{M}|/|\vec{H}|)$,
Eqs.~(\ref{eq:P},\ref{eq:M}) imply that the negative values
of $\overleftrightarrow{\varepsilon_{r}}$ and $\overleftrightarrow{\mu_{r}}$
occur necessarily around the molecular resonant transition frequencies
$\Delta_{k\sigma}$. Furthermore, the responses exist only if the
corresponding transitions are allowed by both \textit{\emph{electric}}
and magnetic dipole interactions with fields. As for the M\"{o}bius
molecular ring, it is particularly interesting that the inter-band
transition $\left|k,\uparrow\right\rangle \rightleftharpoons\left|k^{\prime},\downarrow\right\rangle $
could be allowed both electrically and magnetically by the dipole
couplings. Besides, the typical value for resonance integral is around
the order of a few eV \cite{Greenwood72,Hawkea09}, already in the region of visible
frequency. Therefore, simultaneously negative permittivity and permeability
in the visible frequency regime might be realized in a medium containing
molecules with the M\"{o}bius topology.

\begin{figure}
\begin{centering}
\includegraphics[scale=0.24]{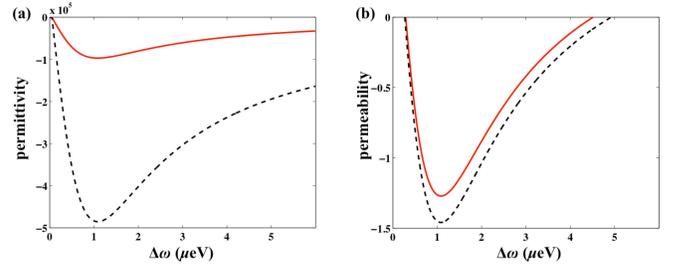}
\par\end{centering}

\caption{(color online) Relative permittivity $\protect\overleftrightarrow{\varepsilon_{r}}$
and permeability $\protect\overleftrightarrow{\mu_{r}}$ as a function
of the detuning $\Delta\omega=\omega-\Delta_{0,\uparrow}$ between
the frequencies of incident field $\omega$ and transition $\left|0,\downarrow\right\rangle \rightleftharpoons\left|0,\uparrow\right\rangle $.
Only negative diagonal elements of $\protect\overleftrightarrow{\varepsilon_{r}}$
and $\protect\overleftrightarrow{\mu_{r}}$ in their corresponding
principal-axis coordinate systems are shown. Here we adopt following parameters $V=\xi=3.6$eV \cite{Greenwood72}, $W=0.077$nm \cite{Silbey04}, $R=NW/\pi$, $\gamma^{-1}=4$ns \cite{Tokuji09}.\label{fig:Permit-Permea}}
\end{figure}

To quantitatively investigate this possibility, $\overleftrightarrow{\varepsilon_{r}}$
is explicitly calculated for the M\"{o}bius ring by using the matrix
elements $\vec{d}_{k\sigma,g}$. For positive $V$ and $\xi$, the
ground state of $H$ is $\left|0,\downarrow\right\rangle $ and thus
$E_{g}=E_{0\downarrow}$. When $\omega$ is near the inter-band transition
frequency $\Delta_{0\uparrow}$ and for sufficiently long lifetime
$\tau$, $\overleftrightarrow{\varepsilon_{r}}$ is simplified from
Eq.~(\ref{eq:P}) by only including the summation term whose denominator
contains a transition frequency that is equal to $\Delta_{0\uparrow}$. With this
approximation, the relative permittivity is simplified in
the molecular coordinate system as
\begin{equation}
\overleftrightarrow{\varepsilon_{r}}(\omega)=\left[\begin{array}{ccc}
1-\eta^{\prime}(\omega) & 0 & 0\\
0 & 1-\eta^{\prime}(\omega) & -2\eta^{\prime}(\omega)\\
0 & -2\eta^{\prime}(\omega) & 1-4\eta^{\prime}(\omega)
\end{array}\right],\label{ep}
\end{equation}
where
\begin{equation}
\eta(\omega)=\frac{1}{8\hbar\varepsilon_{0}\upsilon_{0}}\frac{e^{2}W^{2}}{\omega-2V-2\xi(1-\cos\frac{\delta}{2})+i\gamma},
\label{eta complex}
\end{equation}
and $\eta^{\prime}$ ($\eta^{\prime\prime}$) is real (imaginary)
part of $\eta$. The permittivity tensor is not diagonal in the molecular
coordinate system because of the non-zero off-diagonal term $\varepsilon_{r}^{(yz)}$.
Since the tensor is real-symmetric, $\overleftrightarrow{\varepsilon_{r}}$
is diagonalized in the principal-axis coordinate system by a rotation
around the $x$ axis. This yields
\begin{equation}
\varepsilon_{1}=1-5\eta^{\prime},
\end{equation}
$1-\eta^{\prime}$ and $1$ for relative permittivity along three
principal axes, respectively. In Fig.~\ref{fig:Permit-Permea}(a), the relative permittivity is numerically demonstrated with the following parameters, $V=\xi=3.6$eV \cite{Greenwood72}, $W=0.077$nm \cite{Silbey04}, $\pi R=NW$, $\gamma^{-1}=4$ns \cite{Tokuji09}, which are obtained by fitting the spectra in experiments.

On the other hand, the tensor of the relative permeability $\overleftrightarrow{\mu_{r}}$
is represented by
\begin{equation}
\overleftrightarrow{\mu_{r}}(\omega)\!=\!\left[\begin{array}{ccc}
1-\alpha^{2}\eta^{\prime} & 0 & 0\\
0 & 1-\alpha^{2}\eta^{\prime} & -2\alpha\beta\eta^{\prime}\\
0 & -2\alpha\beta\eta^{\prime} & 1-4\beta^{2}\eta^{\prime}
\end{array}\right],\label{eq:mur}
\end{equation}
where
\begin{eqnarray}
\alpha & = & \frac{R}{\hbar c}[V+\xi(\cos\delta-\cos\frac{\delta}{2})],\label{eq:Alpha}\\
\beta & = & 2\frac{R\xi}{\hbar c}\sin^{2}\frac{\delta}{2}\cos\frac{\delta}{2}.
\end{eqnarray}
The permeability tensor is not diagonal in the molecular coordinate
system either. Because $\alpha$ and $\beta$ are generally different
from 1, $\overleftrightarrow{\mu_{r}}$ and $\overleftrightarrow{\varepsilon_{r}}$
cannot be diagonalized by the same rotating transformation. In other
words, the principal axes in which $\overleftrightarrow{\mu_{r}}$
is diagonal do not coincide with the principal axes of $\overleftrightarrow{\varepsilon_{r}}$.
This yields
\begin{equation}
\mu_{1}=1-\left(\alpha^{2}+4\beta^{2}\right)\eta^{\prime}, \label{mu1}
\end{equation}
$1-\alpha^{2}\eta^{\prime}$, and $1$ for relative permeability along
three principal axes, cf. Fig.~\ref{fig:Permit-Permea}(b), respectively.

\section{Negative Refraction at Medium Interface}

\label{sec:NegRefMedInterface}

The M\"{o}bius medium is anisotropic in the sense that the relative
permittivity $\overleftrightarrow{\varepsilon_{r}}$ and permeability
$\overleftrightarrow{\mu_{r}}$ are tensors rather than scalars. Therefore,
apart from demonstrating simultaneously $\overleftrightarrow{\mu_{r}}<0$
and $\overleftrightarrow{\varepsilon_{r}}<0$ in some frequency region,
the more concrete way to show the existence of negative refraction
is to investigate the behavior of refracted electromagnetic waves at
the medium interface. Here, we investigate the reflection and refraction
at a planar interface between the M\"{o}bius medium and the air for
two specific incident configurations. The results show that the medium
is ``left-handed'' for \emph{E-polarized} propagating mode \cite{Landau95}.

The behavior of a propagating wave inside a medium is captured by
its phase and group velocities. For example, when the negative refraction
was first introduced in Veselago's seminal paper \cite{Veselago68},
a ``left-handed'' material was described as a medium in which the
electromagnetic wave propagates with the opposite phase velocity with
respect to the group velocity. In media where the permittivity and
permeability tensors are symmetric, the group velocity is along the
same direction as the Poynting vector \cite{Landau95}. In such case,
the reversal of the phase velocity to the Poynting vector has been
applied as a criterion for ``left-handed'' materials: $\vec{k}_{t}\cdot\vec{S}_{t}<0$
\cite{Hu02}, where $\vec{k}_{t}$ and $\vec{S}_{t}$ are respectively
the wave vector and Poynting vector of refracted field.
Besides this, if we consider the refraction from the medium interface,
the causality requires that the normal component of its Poynting vector
is in the same direction as that for the incident light \cite{Tao12}.
In summary, we adopt the following three criteria as characteristics
of a ``left-handed'' medium: (i) The wave vector $\vec{k}_{t}$
in the medium is a real-valued vector; (ii) The wave vector $\vec{k}_{t}$
and Poynting vector $\vec{S}_{t}$ in the medium satisfy $\vec{k}_{t}\cdot\vec{S}_{t}<0$;
(iii) The normal component of the Poynting vector remains the same
sign across the medium interface, e.g., if interface normal is $\hat{e}_{z}$
then $S_{iz}S_{tz}>0$, where $\vec{S}_{i}$ is the Poynting vector
of the incident light.

\begin{figure}
\begin{centering}
\includegraphics[scale=0.24]{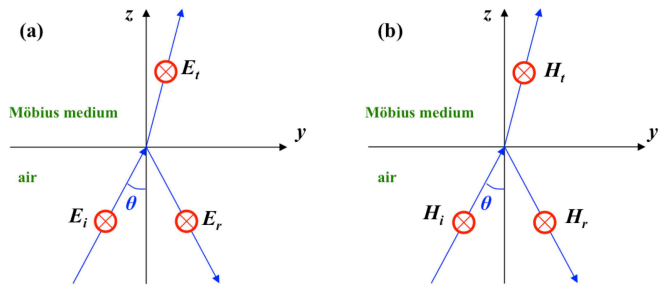}
\par\end{centering}

\caption{(color online) Schematic plots of the reflection and the refraction
at the medium interface $z=0$. (a) The electric field and
(b) the magnetic field of the refracted light is polarized along the $x$-direction, respectively.\label{fig:Schematic-plots-of}}
\end{figure}

Let us consider a linearly polarized monochromatic light which is incident
from the air onto the interface of M\"{o}bius medium as illustrated
in Fig.~\ref{fig:Schematic-plots-of}. Let $z=0$ be the medium interface
and suppose that the incident plane is the $y$-$z$ plane and $\theta$
is the angle between the incident wave vector and the $z$ axis. We
consider two independent incident configurations respectively:

\begin{equation}
\vec{E}_{i}=E_{0}\hat{e}_{x}e^{i(k_{iy}y+k_{iz}z-\omega t)},\mbox{ (\emph{E-polarized})}\label{E-polarized}
\end{equation}
and
\begin{equation}
\vec{H}_{i}=H_{0}\hat{e}_{x}e^{i(k_{iy}y+k_{iz}z-\omega t)},\mbox{ (\emph{H-polarized})}\label{H-polarized}
\end{equation}
where $\vec{k}_{i}=k_{iy}\hat{e}_{y}+k_{iz}\hat{e}_{z}$ is the wave
vector of incident light. These two configurations are known as \emph{E-polarized}
and \emph{H-polarized} respectively, as the electric and magnetic
fields of the refracted light are perpendicular to the refracted wave
vector \cite{Landau95}, respectively.

\subsection{$E$-polarized Incident Configuration}

For the specific incident configuration given by Eq.~(\ref{E-polarized}),
the three criteria for a ``left-handed'' medium could be checked
one by one. It follows from the boundary conditions derived
from the Maxwell's equations that \cite{Jackson99}
\begin{eqnarray}
\hat{e}_{z}\times(\vec{E}_{i}+\vec{E}_{r}-\vec{E}_{t}) & = & 0,\nonumber \\
\hat{e}_{z}\times(\vec{H}_{i}+\vec{H}_{r}-\vec{H}_{t}) & = & 0,\label{BC for H}\\
k_{iy}=k_{ty} & = & k_{i}\sin\theta,\nonumber
\end{eqnarray}
where the subscript $r$ denotes the reflected field back to the air.
In accordance with above boundary conditions, the electric fields
of the reflected and refracted lights could be written as
\begin{eqnarray}
\vec{E}_{r} & = & rE_{0}\hat{e}_{x}e^{i(k_{iy}y-k_{iz}z-\omega t)},\mbox{ }\\
\vec{E}_{t} & = & tE_{0}\hat{e}_{x}e^{i(k_{iy}y+k_{tz}z-\omega t)},\label{Et E-polarized}
\end{eqnarray}
where we also use $r$ and $t$ to represent the reflection and refraction
coefficients respectively, and $1+r=t$. To determine the refracted
wave vector, we combine the two Maxwell's equations $\vec{k}_{t}\times\vec{E}_{t}=\omega\mu_{0}\overleftrightarrow{\mu_{r}}\vec{H}_{t}$
and $\vec{k}_{t}\times\vec{H}_{t}=-\omega\varepsilon_{0}\overleftrightarrow{\varepsilon_{r}}\vec{E}_{t}$:
by multiplying both hand sides of the first equation with $\overleftrightarrow{\mu_{r}}{}^{-1}$
and then inserting it to the second equation, the following equation could be derived, i.e.
\begin{equation}
\vec{k}_{t}\times[(\overleftrightarrow{\mu_{r}})^{-1}(\vec{k}_{t}\times\vec{E}_{t})]=-\omega^{2}\mu_{0}\varepsilon_{0}\overleftrightarrow{\varepsilon_{r}}\vec{E}_{t}.\label{Maxwell's eq E-polarized}
\end{equation}
Since the equation is homogeneous in $E_{tx}$, $E_{ty}$ and $E_{tz}$,
the necessary condition for non-vanishing refracted light is that
the determinant of its coefficient matrix is zero.
Combined with $\vec{E}_{t}=E_{tx}\hat{e}_{x}$ and $k_{tx}=0$,
this necessary condition could be explicitly written as

\begin{figure*}
\centering{}\includegraphics[scale=0.4]{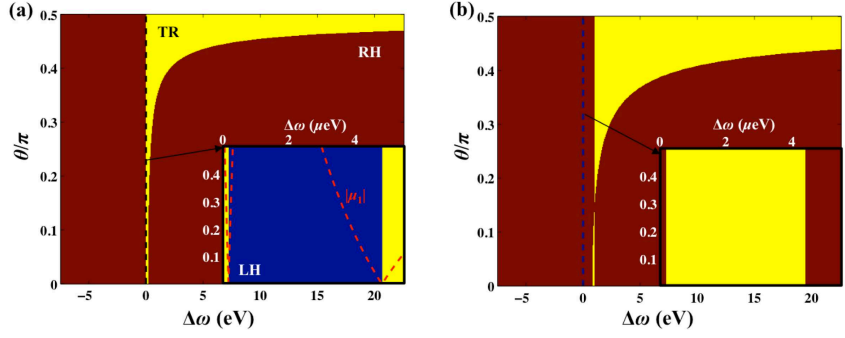} \caption{\label{Fig.S2}(color online) ``Phase diagram'' of left-handedness
in the $\theta$-$\omega$ plane for (a) \textit{E-polarized} incident
field, (b) \emph{H-polarized} incident field. Red and yellow regions
indicate respectively the parameter regions where the refracted wave
is ``right-handed'' (RH) and the incident field is totally reflected
(TR) back to the air. The refracted field is ``left-handed'' (LH)
in the blue region. The inserts show a magnified part in the main
plot where $\omega$ is close to the lowest inter-band transition.
The red dashed curve shows $\vert\mu_{1}\vert$ as a function of $\omega$ near
the inter-band transition. Here we use the same parameters as in Fig.~\ref{fig:Permit-Permea}.}
\end{figure*}

\begin{equation}
\frac{\omega^{2}}{c^{2}}\varepsilon_{r}^{(xx)}-\frac{1}{\mu_{1}}(k_{iy}^{2}\mu_{r}^{(xx)}+2\mu_{r}^{(yz)}k_{iy}k_{tz}+\mu_{r}^{(zz)}k_{tz}^{2})=0.\label{Fresnel equation}
\end{equation}
Inserting Eqs.~(\ref{ep},\ref{eq:mur}) into the above equation yields
\begin{eqnarray}
k_{tz}\!\!\! & = & \!\!\!\{\pm k_{i}\sqrt{\mu_{1}\left[\left(1-\eta^{\prime}\right)\left(1-4\beta^{2}\eta^{\prime}\right)-\sin^{2}\theta\right]}\nonumber \\
\!\!\! &  & \!\!\!-\mu_{r}^{(yz)}k_{iy}\}/\mu_{r}^{(zz)}.\label{kz E-polarized}
\end{eqnarray}

The time-averaged Poynting vector of refracted light $\vec{S}_{t}=\mathrm{Re}\{\vec{E}_{t}\times\vec{H}_{t}^{*}\}/2$
reads
\begin{eqnarray}
\vec{S}_{t} & = & S_{ty}\hat{e}_{y}+S_{tz}\hat{e}_{z},\label{Poynting E-polarized}
\end{eqnarray}
where
\begin{eqnarray}
S_{ty} & = & \frac{E_{0}^{2}t^{2}}{2\omega\mu_{0}\mu_{1}}\left(k_{tz}\mu_{r}^{(yz)}+k_{iy}\mu_{r}^{(xx)}\right),\label{eq:Poynting E-polarized-y}\\
S_{tz} & = & \frac{E_{0}^{2}t^{2}}{2\omega\mu_{0}\mu_{1}}\left(k_{iy}\mu_{r}^{(yz)}+k_{tz}\mu_{r}^{(zz)}\right).\label{eq:Poynting E-polarized-z}
\end{eqnarray}

According to Eqs.~(\ref{eq:mur},\ref{kz E-polarized},\ref{Poynting E-polarized}),
for the \emph{E-polarized} configuration, the three criteria for ``left-handed''
medium are specified as $\vec{k}_{t}\cdot\vec{S}_{t}<0,$ and
\begin{eqnarray}
\left(1-\eta^{\prime}\right)\left(1-4\beta^{2}\eta^{\prime}\right)\!\! & \geq & \!\!\sin^{2}\theta,\mbox{ if }\mu_{1}>0,\\
\left(1-\eta^{\prime}\right)\left(1-4\beta^{2}\eta^{\prime}\right)\!\! & \leq & \!\!\sin^{2}\theta,\mbox{ if }\mu_{1}<0,
\end{eqnarray}
and
\begin{equation}
S_{tz}=\frac{E_{0}^{2}t^{2}}{2\omega\mu_{0}\mu_{1}}[-2\alpha\beta\eta^{\prime}k_{iy}+(1-4\beta^{2}\eta^{\prime})k_{tz}]>0.
\end{equation}

Figure~\ref{Fig.S2}(a) shows the ``phase diagram'' in the $\theta$-$\omega$
plane as determined by the above three criteria. Obviously, there
is a frequency window in which the applied electromagnetic fields
can be negatively refracted, while in most regions of the $\theta$-$\omega$
plane the refracted light are ``right-handed''. Besides these, there
is also a considerable region where the light will be totally reflected.
The inset of Fig.~\ref{Fig.S2}(a) also shows $\mu_{1}$ near the
inter-band transition frequency of the individual M\"{o}bius molecule.
A comparison shows that there is a correspondence between the ``left-handed''
phase boundary and the zeros of $\mu_{1}$, which will be illustrated
later.

To quantitatively understand the ``phase diagram'', we rely on the
geometry relation between the Poynting vector and the wave vector
surface \cite{Landau95}. The wave vector surface is defined by solutions
to Eq.~(\ref{Maxwell's eq E-polarized}) in the $(k_{tx},k_{ty},k_{tz})$
space (or equivalently in the $(n_{tx},n_{ty},n_{tz})$ space by $\vec{n}_{t}\equiv\vec{k}_{t}/\omega\sqrt{\mu_{0}\varepsilon_{0}}$).
This geometry relation then asserts: in the medium with \textit{symmetric}
$\overleftrightarrow{\varepsilon_{r}}$ and $\overleftrightarrow{\mu_{r}}$,
the Poynting vector for a given $\vec{k}$ is either parallel or anti-parallel
to the normal vector of the wave vector surface, as proven in Appendix~\ref{Appendix5}.

\begin{figure}
\begin{centering}
\includegraphics[scale=0.3]{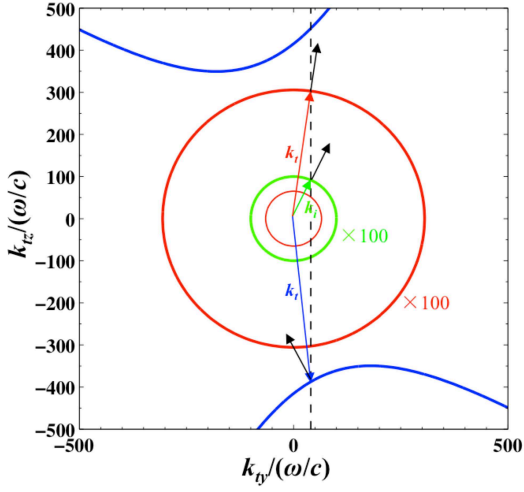}
\par\end{centering}

\caption{(color online) Cross sections of the wave vector surfaces of the refracted
fields at different incident frequencies: $\omega\ll\Delta_{0\uparrow}$
(thick red, magnified by $100$), $\omega\gtrsim\Delta_{0\uparrow}$
(thick blue), $\omega\gg\Delta_{0\uparrow}$ (thin red). The green
circle, magnified by $100$, depicts the wave vector surface of the
incident field (\emph{E-polarized}). In each case, the refracted wave
vector $\vec{k}_{t}$ and the direction of corresponding Poynting
vector $\vec{S}_{t}$ (black arrow) are explicitly shown, while the
wave vector of the incident field $\vec{k}_{i}=\omega\sqrt{\mu_{0}\omega_{0}}(\sin\theta\hat{e}_{y}+\cos\theta\hat{e}_{z})$
is depicted as the green arrow. The black dashed line represents the
constraint from the boundary condition.  Here we use the same parameters as in Fig.~\ref{fig:Permit-Permea}. \label{fig: wave vector surface}}
\end{figure}

The shape of the wave vector surface, as shown in Fig.~\ref{fig: wave vector surface},
strongly depends on the incident frequency. And those different shapes
will in turn give rise to contrasting propagation properties of the
refracted fields. Let us first consider an incident field with frequency
$\omega\ll\Delta_{0\uparrow}$. It follows from Eq.~(\ref{eta complex})
that in this case $\eta'<0$ and $\vert\eta'\vert$ is of the order of unity, while $\alpha\ll1$
and $\beta\ll1$. Therefore, after ignoring the second order terms of
$\alpha$ and $\beta$ in Eq.~(\ref{Fresnel equation}),
the wave vector surface is nearly of circle shape, i.e.,
\begin{equation}
n_{ty}^{2}+n_{tz}^{2}\approx1-\eta^{\prime}.\label{wave vector surface smaller than transition freq}
\end{equation}
Obviously, a real solution of $k_{tz}$ is admitted for all incident
angles, cf. Fig.~\ref{fig: wave vector surface}. Since $\mu_{r}^{(yz)}\ll1$
and meanwhile $\mu_{1}$, $\mu_{r}^{(xx)}$, and $\mu_{r}^{(zz)}$
are of order one when $\omega\ll\Delta_{0\uparrow}$, it follows from
Eqs.~(\ref{Poynting E-polarized}-\ref{eq:Poynting E-polarized-z})
that the Poynting vector $\vec{S}_{t}$ is approximately along the
direction of $\vec{k}_{t}$. Because $\vec{k}_{t}\cdot\vec{S}_{t}<0$ is violated,
the refracted field is ``right-handed''.

When the frequency of the incident field is increased to just above
the lowest inter-band transition frequency $\omega\gtrsim\Delta_{0\uparrow}$,
since $\eta^{\prime}\gg1$, $\mu_{1}$ in Eq.~(\ref{mu1})
could be negative. By linearly combining $n_{ty}$ and $n_{tz}$
to eliminate the cross term in Eq.~(\ref{Fresnel equation}),
the wave vector surface is a hyperboloid, i.e.,
\begin{equation}
\frac{\tilde{n}_{tz}^{2}}{(1-\eta^{\prime})\mu_{1}}-\frac{\tilde{n}_{ty}^{2}}{\eta^{\prime}-1}=1,\label{wave vector surface NR regime}
\end{equation}
where $\tilde{n}_{tz}\!\!=\!\!n_{tz}\sin\phi+n_{ty}\cos\phi$, $\tilde{n}_{ty}\!\!=\!\!n_{tz}\cos\phi-n_{ty}\sin\phi$,
and the mixing angle $\phi\!\!=\!\!\tan^{-1}[-4\alpha\beta/(\beta^{2}-\alpha^{2})]/2$.
Similar non-closed wave vector surface has been suggested in uniaxial
left-handed materials \cite{Shen05}. As $\mu_{1}<0$, Eq.~(\ref{Poynting E-polarized})
implies that the Poynting vector is opposite to the normal vector
of the wave vector surface and thus $\vec{k}_{t}\cdot\vec{S}_{t}<0$,
showing that the wave propagating in the M\"{o}bius medium is now
``left-handed''. Furthermore, in order to fulfill the requirement for causality, the
refracted wave vector should be on the lower branch, cf. Fig.~\ref{fig: wave vector surface}.

Equation~(\ref{wave vector surface NR regime}) also indicates the
frequency region where ``left-handed'' propagating wave is allowed
in the M\"{o}bius medium. Since $\alpha$ and $\beta$ are small
quantities, the sign of $\mu_{1}$ will change along with the increase
of $\eta^{\prime}$. This sign change will make the shape of the wave
vector surface deform from a hyperboloid to a spheroid. However, because
$1-\eta^{\prime}$ is still negative and $\mu_{1}>0$, Eq.~(\ref{wave vector surface NR regime})
will not have solution for real $n_{tz}$. An imaginary $n_{tz}$
indicates that the wave can not propagate inside the medium and the
incident wave is totally reflected. Therefore, the frequency region
for the propagating ``left-handed'' \textit{E-polarized} wave in
the M\"{o}bius medium is determined by the sign change of $\mu_{1}$,
cf. Fig.~\ref{Fig.S2}(a), and the resulting bandwidth, i.e., width
of the frequency region supporting negative refraction, is $\mathcal{B}=\vert\omega_{1}-\omega_{2}\vert$,
where $\omega_{j}$s are the two solutions to $\mu_{1}(\omega_{j})=0$.

If we further tune the frequency $\omega$ to be far bigger than $\Delta_{0\uparrow}$,
$\eta^{\prime}$ approaches zero, i.e. $\eta^{\prime}\rightarrow0^{+}$.
Equation~(\ref{wave vector surface smaller than transition freq})
is again valid in this case, but the radius of the wave vector surface
for the refracted field is smaller than that of the incident field.
However, $\vec{k}_{t}\cdot\vec{S}_{t}>0$ suggests that the refracted light
is also ``right-handed'' in this case.
Furthermore, the total reflection of the incident light will happened
as there is no real solution for $n_{tz}$ at large incident angle.

\subsection{$H$-polarized Incident Configuration}

The analysis for the \emph{H-polarized} case is similar to that for the \emph{E-polarized} case.
In accordance with Eqs.~(\ref{BC for H},\ref{H-polarized}),
the reflected and refracted magnetic fields are written as, cf. Fig.~\ref{fig:Schematic-plots-of}(b)
\begin{eqnarray}
\vec{H}_{r} & = & rH_{0}\hat{e}_{x}e^{i(k_{iy}y-k_{iz}z-\omega t)},\\
\vec{H}_{t} & = & tH_{0}\hat{e}_{x}e^{i(k_{iy}y+k_{tz}z-\omega t)}.
\end{eqnarray}
The equation for determining the refracted wave vector component $k_{tz}$
is similar to Eq.~(\ref{Maxwell's eq E-polarized}), i.e.,
\begin{equation}
\vec{k}_{t}\times\left[(\overleftrightarrow{\varepsilon_{r}})^{-1}\left(\vec{k}_{t}\times\vec{H}_{t}\right)\right]=-\omega^{2}\mu_{0}\varepsilon_{0}\overleftrightarrow{\mu_{r}}\vec{H}_{t}.
\end{equation}
By solving this equation, we obtain the refracted wave vector as
\begin{eqnarray}
k_{tz}\!\!\! & = & \!\!\!\{\pm k_{i}\sqrt{\varepsilon_{1}[(1-\eta^{\prime}\alpha^{2})(1-4\eta^{\prime})-\sin^{2}\theta]}\nonumber \\
\!\!\! &  & \!\!\!-\varepsilon_{r}^{(yz)}k_{iy}\}/\varepsilon_{r}^{(zz)}.\label{P-1}
\end{eqnarray}

The time-averaged Poynting vector for the refracted light is
$\vec{S}_{t}=S_{ty}\hat{e}_{y}+S_{tz}\hat{e}_{z}$, with
\begin{equation}
S_{ty}=\frac{H_{0}^{2}t^{2}}{2\omega\varepsilon_{0}\varepsilon_{1}}(k_{tz}\varepsilon_{r}^{(yz)}+k_{iy}\varepsilon_{r}^{(xx)}),
\end{equation}
and
\begin{equation}
S_{tz}=\frac{H_{0}^{2}t^{2}}{2\omega\varepsilon_{0}\varepsilon_{1}}(k_{iy}\varepsilon_{r}^{(yz)}+k_{tz}\varepsilon_{r}^{(zz)}).
\end{equation}

The three criteria for ``left-handed'' medium are then summarized
as $\vec{k}_{t}\cdot\vec{S}_{t}<0$, and
\begin{eqnarray}
\left(1-\eta^{\prime}\alpha^{2}\right)\left(1-4\eta^{\prime}\right) & \geq & \sin^{2}\theta,\mbox{ if }5\eta^{\prime}<1,\\
\left(1-\eta^{\prime}\alpha^{2}\right)\left(1-4\eta^{\prime}\right) & \leq & \sin^{2}\theta,\mbox{ if }5\eta^{\prime}>1,
\end{eqnarray}
and
\begin{equation}
(1-5\eta^{\prime})[-2\eta^{\prime}k_{iy}+(1-4\eta^{\prime})k_{tz}]>0.
\end{equation}

The ``phase diagram'' in the $\theta$-$\omega$ plane for the \emph{H}-\emph{polarized}
case is deduced according to above three criteria, as shown in Fig.~\ref{Fig.S2}(b).
Similarly to the \textit{E-polarized} case, there are regions in the
$\theta$-$\omega$ plane where the refracted propagating field is ``right-handed''
and regions where the incident field is totally reflected. However,
in contrast to the \textit{E-polarized} case, there are no frequency
regions in the \textit{H-polarized} case such that the propagating refracted field could be ``left-handed''.

\subsection{Discussions}

In the previous subsections, we have discussed the negative refraction regardless of loss.
Generally speaking, there will be loss in the medium due to couplings of the molecules to the bath.
The condition for negative refraction in the presence of loss is slightly
different from that without loss \cite{McCall02,Depine04}. Because
of the loss effect, the imaginary part $\eta^{\prime\prime}$ should
also be considered in the relative permittivity and permeability.
This results in the replacement of $\eta^{\prime}$ by $\eta=\eta^{\prime}+i\eta^{\prime\prime}$
in Eqs.~(\ref{ep},\ref{eq:mur}). Secondly, the refracted wave vector
as determined from the Maxwell's equations is now generally complex,
and the phase velocity is along the real part of the refracted wave
vector \cite{McCall02,Depine04}. Through considering the normal incidence of
an \emph{E-polarized} field onto the medium interface,
we can prove that the frequency region of negative refraction
is not qualitatively changed when the loss is taken into consideration.

On the other hand, as depicted in Eq.~(\ref{eta complex}),
the negative refraction will apparently disappear if the decay rate of excited stats is sufficiently large.
A limitation on the lifetime of the excited state for achieving negative
refraction could be deduced from the bandwidth $\mathcal{B}=|\omega_{1}-\omega_{2}|$
according to zeros of $\mu_{1}(\omega)$, i.e.
\begin{eqnarray}
\mathrm{\mathcal{B}} & = & \mathrm{Re}\sqrt{\left[\frac{e^{2}\left(\alpha^{2}+4\beta^{2}\right)W^{2}}{8\varepsilon_{0}\upsilon_{0}\hbar}\right]^{2}-4\gamma^{2}}.\label{bandwidth lossless}
\end{eqnarray}
This indicates a restriction on the excited state lifetime
\begin{equation}
\tau_{c}=\frac{16\varepsilon_{0}\upsilon_{0}\hbar}{e^{2}\left(\alpha^{2}+4\beta^{2}\right)W^{2}},
\end{equation}
above which the negative refraction from the M\"{o}bius medium is
expectable. Numerical verification has been performed and shows that
the bandwidth is not qualitatively changed with the inclusion of the
loss effect. In experiments, it is possible to synthesize
a M\"{o}bius ring of carbon atoms with $N=12$ and radius $0.29$nm
\cite{Yoneda14,Balzani08}. And it was theoretically predicted that
M\"{o}bius molecules with more than $60$ atoms are as stable as H\"{u}ckel molecules \cite{Estrada12}.
By taking the value $3.6$eV for $V$ and $\xi$ \cite{Greenwood72},
one finds that an excited state lifetime
of $0.51$ns is enough for observing negative refraction. Since the
excited-state lifetime of M\"{o}bius systems can reach as long as
$350$ns \cite{Tokuji09,Yoon09}, it is reasonable to expect a medium
with M\"{o}bius molecules as a potential material to show the negative
refraction.

Although in our calculation only one electron is considered, our result
is consistent with the more realistic case when all $\pi$ electrons
from all atoms in the M\"{o}bius molecule are taken into account.
In that case, when the spin degree of freedom is considered, two electrons
with different spin states can stay in the same energy eigenstate
$\vert k,\sigma\rangle$. For the ground state of the total system
including all electrons, all the states of the lower energy band will
be filled. Theoretically, there could be negative refraction around
$4N$ possible transition frequencies if the excited-state lifetime
is sufficiently long, as implied by Eqs.~(\ref{eq:Dk0},\ref{eq:Mk0},\ref{bandwidth lossless}).
The inter-band transition in our calculation is just the special case
of $4N$ possible transitions. In this sense, our simplified calculation
clearly illustrates the key factors for demonstrating negative refraction
in M\"{o}bius molecules.

Furthermore, the Hamiltonian describing a molecule interacting with electromagnetic field
is approximated as dipole interaction in our calculation, cf. Eqs.~(\ref{eq:He},\ref{eq:Hb}).
Generally speaking, there are multi-pole contributions to Coulomb interaction
between molecule and electromagnetic field. As long as the molecule is small,
the dipole approximation is valid and has been frequently used
in the investigation of metamaterials \cite{Jackson99,Thommen06}.
On the other hand, for the sake of simplicity, the inter-molecular interaction
has been neglected in our calculation. As shown in the Appendix \ref{Appendix_Add},
based on the Lorentz local field theory, both the central frequency and bandwidth
of negative refraction have not been substantially modified when the interaction between molecules
is taken into account. As a result, by modeling M\"{o}bius molecules
as non-interacting particles, the key factors influencing negative refraction
can be captured.

The M\"{o}bius molecule discussed in this paper is termed an equilateral M\"{o}bius strip
as the twist density is the same everywhere. In Ref.~\cite{Walba93}, it was reported that
two conformations of tetrahydroxymethylethylene M\"{o}bius molecule with chemical formula
C$_{42}$H$_{72}$O$_{18}$ have been synthesized. Although the equilateral M\"{o}bius molecule
has not yet been synthesized, it was predicted that it could be achievable~\cite{Walba93}.
On the other hand, we notice that the H\"{u}ckel model with empirical parameters
has been successfully applied to describing experimental data of more than 60 organic molecules
with maximum deviation no more than 15\%~\cite{Hawkea09,Greenwood72,Silbey04}.
Therefore, it is reasonable to expect that the theoretical predictions
for M\"{o}bius molecules could also be observed in the future experiments.

\section{Conclusion}

\label{Conclusion}

We have explored the M\"{o}bius molecular ring as a potential candidate
for negative refraction. The previous investigations with the functional
atoms or molecules rely on the conceptual analogy of the split-ring
resonator for magnetic response \cite{Shen14}, while for M\"{o}bius
ring this is induced by its non-trivial boundary condition. Our results
demonstrate that engineering on the topology is beneficial in realizing
the high frequency magnetic response at the molecular level. This
finding opens up an alternative approach to design molecular negative
index materials, which is promising in achieving 3D bulk negative
refraction at the visible wavelength.

We further remark that our proposal is complementary to the previous
experimental investigation \cite{Chang10}, where the classical metamaterial
was fabricated with M\"{o}bius topology in a ``top-down'' fashion.
In order to induce the magnetic response, their element is still based
on the configuration of the split-ring resonator. Moreover, due to
quantum effect, our architecture is two order smaller in size than
theirs.

This work was supported by the National Natural Science Foundation
of China (Grant No.~11121403 and No.~11505007), the National 973-program
(Grant No.~2012CB922104 and No.~2014CB921403), and the Youth Scholars
Program of Beijing Normal University (Grant No.~2014NT28), and the
Open Research Fund Program of the State Key Laboratory of Low-Dimensional
Quantum Physics, Tsinghua University Grant No.~KF201502.

\appendix

\section{Absence of Magnetic Dipole Transition in a Perfect Ring}
\label{Appendix1}

Consider a molecular ring formed by $N$ identical atoms with the
nearest neighbour hopping strength $\xi$ and site energy $\epsilon$.
The single electron Hamiltonian is written as
\begin{equation}
H=\sum_{j=0}^{N-1}\left[\epsilon a_{j}^{\dagger}a_{j}-\left(\xi a_{j}^{\dagger}a_{j+1}+\mathrm{h.c.}\right)\right],\label{perfect ring Hamiltonian}
\end{equation}
where periodical boundary condition $a_{0}=a_{N}$ is assumed. The
operator $a_{j}^{\dagger}$ creates an excitation at the $j$th atom,
which is located at $\vec{R}_{j}=R\cos\varphi_{j}\hat{e}_{x}+R\sin\varphi_{j}\hat{e}_{y}$
and $\varphi_{j}$ is defined in Eq.~(\ref{phi}), $R$ denotes the
radius of the molecular ring.

Based on the bond current formalism in Ref.~\cite{Ceulemans98},
the magnetic dipole operator
$\vec{m}$ reads
\begin{equation}
m_{x}=m_{y}=0,\mbox{ }m_{z}=\frac{i}{2}e\xi R^{2}\sin\delta\sum_{j}a_{j}^{\dagger}a_{j+1}+\mathrm{h.c.}\label{mz perfect ring}
\end{equation}
Equations~(\ref{perfect ring Hamiltonian},\ref{mz perfect ring})
indicate that a perfect ring does not couple to the magnetic field
\cite{Brechtefeld06}. Because $[\vec{m},H]=0$ and the interaction Hamiltonian is proportional
to $\vec{m}$, the interaction does not mix different eigenstates of $H$.
Consequently, the negative refraction is absent in this situation.

Furthermore, we can show that for a common double ring, i.e., a chemical annulene with the periodical boundary condition $a_{0}=a_{N}$ and $b_{0}=b_{N}$, the magnetic-dipole transition
is not at the same frequency as the electric-dipole transition. As a result,
there will not be the negative refraction either.

\section{Diagonalization of M\"{o}bius Hamiltonian}
\label{Appendix2}

Before evaluating matrix elements for dipole operators,
we solve explicitly the energy spectrum and the molecular eigenstates.
Consider a M\"{o}bius molecular ring with $2N$ sites,
which is described by the H\"{u}ckel Hamiltonian $H$ as illustrated in Eq.~(\ref{Hamiltonian}). The Hamiltonian $H$ is in diagonal form when expressed in terms of $\mathbf{C}_{k}$ and
$\mathbf{C}_{k}^{\dagger}$, i.e.,
\begin{equation}
H=\sum_{k}\mathbf{C}_{k}^{\dagger}\mathbf{E}_{k}\mathbf{C}_{k},\label{Mobius Hamiltonian}
\end{equation}
with
\begin{eqnarray}
\mathbf{C}_{k} & = & \left[\begin{array}{c}
d_{k\uparrow}\\
d_{k\downarrow}
\end{array}\right]\mbox{ } =\mbox{ }\frac{1}{\sqrt{N}}\sum_{j=0}^{N-1}e^{ikj}\mathbf{B}_{j},\label{eq.Ck}\\
\mathbf{E}_{k} & = & \left[\begin{array}{cc}
V-2\xi\cos(k-\frac{\delta}{2}) & 0\\
0 & -V-2\xi\cos k
\end{array}\right].
\end{eqnarray}
Therefore, the single-electron molecular eigenstates $|k,\sigma\rangle$
are
\begin{equation}
|k,\uparrow\rangle=d_{k\uparrow}^{\dagger}|0\rangle,\mbox{ }|k,\downarrow\rangle=d_{k\downarrow}^{\dagger}|0\rangle.
\end{equation}
Here, $|0\rangle$ denotes the vacuum state and $k=l\delta$ with
$l=0,1,...,N-1$ and $\delta=2\pi/N$.

It is useful to express $|k,\sigma\rangle$ in terms of the localized
atomic-orbitals for later use. To achieve this one notice that,
according to Ref.~\cite{Zhao09}, $\mathbf{B}_{j}$ is related to
$\mathbf{A}_{j}$ by a local unitary transformation, i.e.
\begin{equation}
\mathbf{B}_{j}=\left[\begin{array}{c}
c_{j+}\\
c_{j-}
\end{array}\right]=\mathbf{U}_{j}\mathbf{A}_{j}\label{eq:B}
\end{equation}
with
\begin{eqnarray}
\mathbf{U}_{j} & = & \frac{1}{\sqrt{2}}\left[\begin{array}{cc}
e^{-i\varphi_{j}/2} & -e^{-i\varphi_{j}/2}\\
1 & 1
\end{array}\right],\label{Local unitary transformation}
\end{eqnarray}
and $j=0,1,...,N-1$. It follows from Eqs.~(\ref{eq.Ck},\ref{eq:B})
that the eigenstates $|k,\sigma\rangle$ can be written in terms of
$\mathbf{A}_{j}=\left[\begin{array}{cc}
a_{j} & b_{j}\end{array}\right]^{\mathrm{T}}$ as
\begin{eqnarray}
\left[\begin{array}{c}
|k,\uparrow\rangle\\
|k,\downarrow\rangle
\end{array}\right] \!\!\!& = &\!\!\! \sum_{j=0}^{N-1}\frac{e^{-ikj}}{\sqrt{N}}\left[\begin{array}{c}
c_{j\uparrow}^{\dagger}\\
c_{j\downarrow}^{\dagger}
\end{array}\right]|0\rangle\nonumber \\
\!\!\! & = & \!\!\! \sum_{j=0}^{N-1}\frac{e^{-ikj}}{\sqrt{2N}}\!\!\!\left[\begin{array}{cc}
e^{i\varphi_{j}/2} & -e^{i\varphi_{j}/2}\\
1 & 1
\end{array}\right]\!\!\!\left[\begin{array}{c}
a_{j}^{\dagger}\\
b_{j}^{\dagger}
\end{array}\right]\!\!\!|0\rangle.
\end{eqnarray}
Furthermore, by noticing that the atomic orbitals $|\phi_{j\pm}\rangle$
are created by acting $a_{j}^{\dagger}$ or $b_{j}^{\dagger}$ on
the vacuum state, the molecular eigenstates are rewritten as
\begin{eqnarray}
|k,\uparrow\rangle & = & \frac{1}{\sqrt{2N}}\sum_{j=0}^{N-1}e^{-i(k-\frac{\delta}{2})j}\left(|\phi_{j+}\rangle-|\phi_{j-}\rangle\right),\label{eq:eigenstate1 Mobius}\\
|k,\downarrow\rangle & = & \frac{1}{\sqrt{2N}}\sum_{j=0}^{N-1}e^{-ikj}\left(|\phi_{j+}\rangle+|\phi_{j-}\rangle\right).\label{eq:eigenstate2 Mobius}
\end{eqnarray}
Especially, if the resonance integrals $V$ and $\xi$ are positive,
the molecular ground state is
\begin{equation}
|g\rangle=|0,\downarrow\rangle=\frac{1}{\sqrt{2N}}\sum_{j=0}^{N-1}\left(|\phi_{j+}\rangle+|\phi_{j-}\rangle\right).
\end{equation}

\section{Matrix Elements of Electric and Magnetic Dipoles}
\label{Appendix3}

The matrix elements of electric dipole operator are
\begin{eqnarray}
\left\langle 0\left|\mathbf{C}_{k}\vec{d}\mathbf{C}_{k}^{\dagger}\right|0\right\rangle \!\!\! & = & \!\!\! -e\frac{W}{4}\left[\left(\hat{e}_{y}+2\hat{e}_{z}\right)\sigma_{x}-\hat{e}_{x}\sigma_{y}\right],\label{EqC1}\\
\left\langle 0\left|\mathbf{C}_{k}\vec{d}\mathbf{C}_{k\pm\delta}^{\dagger}\right|0\right\rangle \!\!\! & = & \!\!\! -e\frac{1}{4}[\left(\hat{e}_{x}\mp i\hat{e}_{y}\right)\left(2R+W\sigma_{y}\right) \nonumber \\
\!\!\! && \!\!\! +2W\hat{e}_{z}\sigma_{\mp}],\\
\left\langle 0\left|\mathbf{C}_{k}\vec{d}\mathbf{C}_{k\pm2\delta}^{\dagger}\right|0\right\rangle \!\!\! & = & \!\!\! -e\frac{W}{4}\left(\mp i\hat{e}_{x}-\hat{e}_{y}\right)\sigma_{\mp},
\end{eqnarray}
where $\vec{d}=-e\vec{r}$ is the electric dipole operator with $-e$
being the electric charge and $\vec{r}$ being the position vector of electron,
$\sigma_{\pm}=(\sigma_{x}\pm i\sigma_{y})/2$ with $\sigma_{\alpha}$
($\alpha=x,y,z$) being Pauli operators in the pseudo spin space spanned by $|k,\sigma\rangle$ and $|k^{\prime},\sigma\rangle$, e.g., for Eq.~(\ref{EqC1})  $\sigma_{z}$ is defined as

\begin{equation}
\sigma_{z}  =  |k,\uparrow\rangle\langle k,\uparrow| -|k,\downarrow\rangle\langle k,\downarrow |,
\end{equation}$\hat{e}_{\alpha}$($\alpha=x,y,z$)
is the unit vector in $\alpha$ direction, $R$ and $4W$ are respectively the radius
and width of the M\"{o}bius molecule.

The matrix elements of magnetic dipole operator $\vec{m}=-ie\vec{r}\times[H,\vec{r}]/2\hbar$
are summarized as follows:
\begin{widetext}
For $|k,\sigma\rangle\rightleftharpoons|k,\sigma^{\prime}\rangle$
transitions
\begin{align}
 & -\left\langle 0\left|\mathbf{C}_{k}\vec{m}\mathbf{C}_{k}^{\dagger}\right|0\right\rangle \nonumber \\
 & =\frac{e}{2\hbar}\left[\begin{array}{cc}
\begin{array}{c}
-\frac{1}{8}\xi\{2W^{2}\left[\cos(k-\delta)-\cos k\right]\hat{e}_{y}+[W^{2}(\cos k\\
-\cos(k-2\delta)-\cos(k-\delta)+\cos(k+\delta))\\
+4R^{2}(\cos(k+\frac{\delta}{2})-\cos(k-\frac{3}{2}\delta))]\hat{e}_{z}\}
\end{array} & \begin{array}{c}
\frac{1}{4}RW\{-\left[V+\xi(\cos(k-\delta)-\cos(k+\frac{\delta}{2}))\right](\hat{e}_{x}-i\hat{e}_{y})\\
-2i\xi\cos\frac{\delta}{4}\left[\cos(k-\frac{5}{4}\delta)-\cos(k+\frac{3}{4}\delta)\right]\hat{e}_{z}\}
\end{array}\\
\begin{array}{c}
\frac{1}{4}RW\{-\left[V+\xi(\cos(k-\delta)-\cos(k+\frac{\delta}{2})\right](\hat{e}_{x}+i\hat{e}_{y})\\
-2i\xi\cos\frac{\delta}{4}\left[\cos(k+\frac{3}{4}\delta)-\cos(k-\frac{5}{4}\delta)\right]\hat{e}_{z}\}
\end{array} & -\frac{1}{2}\xi\sin k\left[W^{2}\sin\frac{\delta}{2}\hat{e}_{y}-(2R^{2}+W^{2}\cos\frac{\delta}{2})\sin\delta\hat{e}_{z}\right]
\end{array}\right],\nonumber \\
\end{align}
for $|k,\sigma\rangle\rightleftharpoons|k\pm\delta,\sigma^{\prime}\rangle$
transitions
\begin{align}
 & -\left\langle 0\left|\mathbf{C}_{k}\vec{m}\mathbf{C}_{k+\delta}^{\dagger}\right|0\right\rangle \nonumber \\
= & \frac{e}{2\hbar}\left[\begin{array}{cc}
\frac{1}{8}W^{2}\xi\left[\cos(k-\delta)-\cos(k+\delta)\right](i\hat{e}_{x}+\hat{e}_{y}-\hat{e}_{z}) & \begin{array}{c}
-\frac{1}{4}RW\{\left[V+\xi\left(\cos k-\cos(k+\frac{\delta}{2})\right)\right](\hat{e}_{x}-i\hat{e}_{y})\\
-2i\xi\cos\frac{\delta}{4}\left[\cos(k-\frac{5}{4}\delta)-\cos(k+\frac{3}{4}\delta)\right]\hat{e}_{z}\}
\end{array}\\
\begin{array}{c}
\frac{1}{4}RW\{\left[V+\xi(\cos(k+\delta)-\cos(k-\frac{\delta}{2}))\right](\hat{e}_{x}-i\hat{e}_{y})\\
-i\xi[\cos(k-\delta)-\cos(k+\delta)-\cos(k+\frac{3}{2}\delta)\\
+\cos(k-\frac{\delta}{2})]\hat{e}_{z}\}
\end{array} & \frac{1}{8}W^{2}\xi\left[\cos(k-\frac{\delta}{2})-\cos(k+\frac{3}{2}\delta)\right](i\hat{e}_{x}+\hat{e}_{y}-\hat{e}_{z})
\end{array}\right],\nonumber \\
\end{align}
and
\begin{align}
 & -\left\langle 0\left|\mathbf{C}_{k}\vec{m}\mathbf{C}_{k-\delta}^{\dagger}\right|0\right\rangle \nonumber \\
= & \frac{e}{2\hbar}\left[\begin{array}{cc}
-\frac{1}{8}W^{2}\xi\left[\cos(k-2\delta)-\cos k\right](i\hat{e}_{x}-\hat{e}_{y}+\hat{e}_{z}) & \begin{array}{c}
\frac{1}{4}RW\{\left[V+\xi(\cos k-\cos(k-\frac{3}{2}\delta))\right](\hat{e}_{x}+i\hat{e}_{y})\\
+i\xi[\cos(k-\frac{3}{2}\delta)+\cos(k-2\delta)-\cos k\\
-\cos(k+\frac{\delta}{2})]\hat{e}_{z}\}
\end{array}\\
-\frac{1}{4}RW\left\{ V+\xi\left[\cos(k-\delta)-\cos(k-\frac{\delta}{2})\right]\right\} (\hat{e}_{x}+i\hat{e}_{y}) & \frac{1}{8}W^{2}\xi\left[\cos(k-\frac{3}{2}\delta)-\cos(k+\frac{\delta}{2})\right](-i\hat{e}_{x}+\hat{e}_{y}-\hat{e}_{z})
\end{array}\right].\nonumber \\
\end{align}
for $|k,\sigma\rangle\rightleftharpoons|k\pm2\delta,\sigma^{\prime}\rangle$
transitions
\begin{align}
 & -\left\langle 0\left|\mathbf{C}_{k}\vec{m}\mathbf{C}_{k+2\delta}^{\dagger}\right|0\right\rangle \nonumber \\
= & \frac{e}{2\hbar}\left[\begin{array}{cc}
\frac{i}{8}W^{2}\xi\left[\cos k-\cos(k+\delta)\right](\hat{e}_{x}-i\hat{e}_{y}) & 0\\
\frac{1}{4}RW\left\{ V+\xi\left[\cos(k+\delta)-\cos(k+\frac{\delta}{2})\right]\right\} (\hat{e}_{x}-i\hat{e}_{y}) & \frac{i}{8}W^{2}\xi\left[\cos(k+\frac{\delta}{2})-\cos(k+\frac{3}{2}\delta)\right](\hat{e}_{x}-i\hat{e}_{y})
\end{array}\right],
\end{align}
and
\begin{align}
 & -\left\langle 0\left|\mathbf{C}_{k}\vec{m}\mathbf{C}_{k-2\delta}^{\dagger}\right|0\right\rangle \nonumber \\
= & \frac{e}{2\hbar}\left[\begin{array}{cc}
-\frac{i}{8}W^{2}\xi\left[\cos(k-2\delta)-\cos(k-\delta)\right](\hat{e}_{x}+i\hat{e}_{y}) & \frac{1}{4}RW\left\{ V+\xi\left[\cos(k-\delta)-\cos(k-\frac{3}{2}\delta)\right]\right\} (\hat{e}_{x}+i\hat{e}_{y})\\
0 & -\frac{i}{8}W^{2}\xi\left[\cos(k-\frac{3}{2}\delta)-\cos(k-\frac{\delta}{2})\right](\hat{e}_{x}+i\hat{e}_{y})
\end{array}\right],
\end{align}
\end{widetext}
where $V$ and $\xi$ are respectively the inter-ring and intra-ring
resonance integrals, $k$ is the momentum of the eigenstate, $\delta=2\pi/N$.

\section{Polarization and Magnetization}
\label{Appendix4}

In order to calculate the polarization of medium, we first consider
the electric dipole $\langle \vec{d}\rangle $ of a single
molecule. According to the linear response theory \cite{Mukamel95}
\begin{equation}
\langle \vec{d}\rangle =\int\frac{d\omega_{1}}{2\pi}S^{(1)}(\omega_{1})\vec{E}(\omega_{1})e^{-i\omega_{1}t},
\end{equation}
where $\vec{E}(\omega_{1})$ is the Fourier transform of the electric field
\begin{equation}
\vec{E}(\omega_{1})=\int_{-\infty}^{\infty}dt\vec{E}(t)e^{i\omega_{1}t},
\end{equation}
and the linear response function in the frequency domain is
\begin{equation}
S^{(1)}(\omega_{1})=-J(\omega_{1})-J^{*}(-\omega_{1}).
\end{equation}
Here, $J(\omega_{1})$ is the dipole-dipole correlation function \cite{Mukamel95}
\begin{equation}
J(\omega_{1})=-i\int_{0}^{\infty}dt\mathrm{Tr}[\vec{d}(t)\vec{d}\rho_{0}]e^{i\omega_{1}t}\label{J_omega}
\end{equation}
with $\rho_{0}=|g\rangle \langle g|$ and
\begin{equation}
\vec{d}(t)=e^{iHt/\hbar}\vec{d}e^{-iHt/\hbar}.
\end{equation}

For a monochromatic continuous driving with frequency $\omega$ and time-independent
envelope $\vec{E}_{0}$, the electric field is
\begin{equation}
\vec{E}(t)=\vec{E}_{0}\cos\omega t,
\end{equation}
and $\langle \vec{d}\rangle $ is rewritten as
\begin{eqnarray}
\langle \vec{d}\rangle  \!\!\!& = &\!\!\! \int\frac{d\omega_{1}}{2\pi}S^{(1)}(\omega_{1})\pi\vec{E}_{0}[\delta(\omega_{1}-\omega)+\delta(\omega_{1}+\omega)]e^{-i\omega_{1}t}\nonumber \\
\!\!\! & = &\!\!\! \frac{1}{2}[S^{(1)}(\omega)e^{-i\omega t}+S^{(1)}(-\omega)e^{i\omega t}]\vec{E}_{0}\nonumber \\
\!\!\! & = &\!\!\! -\frac{1}{2}\{ [J(\omega)+J^{*}(-\omega)]e^{-i\omega t}+[J(-\omega)+J^{*}(\omega)]e^{i\omega t}\} \nonumber \\ && \times \vec{E}_{0}.
\end{eqnarray}
According to Eq.~(\ref{J_omega}), we have
\begin{eqnarray}
J(\omega) & = & -i\int_{0}^{\infty}dt\langle g|\vec{d}(t)\vec{d}|g\rangle \nonumber \\
 & = & -i\sum_{k,\sigma}\mbox{}^{\prime}\int_{0}^{\infty}dte^{i(\omega-\Delta_{k\sigma}+i\gamma)t}\vec{d}_{g,k\sigma}\vec{d}_{k\sigma,g}\nonumber \\
 & = & \sum_{k\sigma}\mbox{}^{\prime}\frac{\vec{d}_{g,k\sigma}\vec{d}_{k\sigma,g}}{\omega-\Delta_{k\sigma}+i\gamma},
\end{eqnarray}
where $\vec{d}_{k\sigma,g}=\langle k,\sigma|\vec{d}|g\rangle $
is matrix element of the electric dipole operator and the decay rate
$\gamma$ (related to the lifetime of the excited states $\tau$ by
$\gamma=1/\tau$) is introduced to phenomenologically describe the
quantum dynamics due to coupling to the bath. The transition frequency
from the ground state to excited state $|k,\sigma\rangle$ is denoted
as $\Delta_{k\sigma}$. The electric dipole is then written as
\begin{widetext}
\begin{eqnarray}
\langle \vec{d}\rangle  & = & -\sum_{k,\sigma}\mbox{}^{\prime}\frac{\vec{d}_{g,k\sigma}\vec{d}_{k\sigma,g}\cdot\vec{E}_{0}}{2\hbar}[(\frac{1}{\omega-\Delta_{k\sigma}+i\gamma}-\frac{1}{\omega+\Delta_{k\sigma}+i\gamma})e^{-i\omega t}+(\frac{1}{\omega-\Delta_{k\sigma}-i\gamma}-\frac{1}{\omega+\Delta_{k\sigma}-i\gamma})e^{i\omega t}]\nonumber \\
 & \approx & -\sum_{k,\sigma}\mbox{}^{\prime}\frac{\vec{d}_{g,k\sigma}\vec{d}_{k\sigma,g}\cdot\vec{E}_{0}}{2\hbar}(\frac{1}{\omega-\Delta_{k\sigma}+i\gamma}e^{-i\omega t}+\frac{1}{\omega-\Delta_{k\sigma}-i\gamma}e^{i\omega t}),
\end{eqnarray}
\end{widetext}
where we have invoked the rotating wave approximation and ignored
the terms with $\omega+\Delta_{k\sigma}$ in the denominator \cite{Mukamel95}.
Re-expressing the above equation in terms of $\vec{E}(t)$ yields
\begin{eqnarray}
\langle \vec{d}\rangle &=& -\mathrm{Re}\sum_{k,\sigma}\mbox{}^{\prime}\frac{\vec{d}_{g,k\sigma}\vec{d}_{k\sigma,g}\cdot\vec{E}(t)}{\hbar(\omega-\Delta_{k\sigma}+i\gamma)} \nonumber\\
&&+\mathrm{Im}\sum_{k,\sigma}\mbox{}^{\prime}\frac{\vec{d}_{g,k\sigma}\vec{d}_{k\sigma,g}\cdot\vec{E}(t-\frac{\pi}{2\omega})}{\hbar(\omega-\Delta_{k\sigma}+i\gamma)}.\label{eq:P1}
\end{eqnarray}
Since the second term on the r.h.s. of Eq.~(\ref{eq:P1}) is related
to absorption and loss, we only keep the real part, i.e.,
\begin{equation}
\langle \vec{d}\rangle =-\mathrm{Re}\sum_{k,\sigma}\mbox{}^{\prime}\frac{\vec{d}_{g,k\sigma}\vec{d}_{k\sigma,g}\cdot\vec{E}(t)}{\hbar(\omega-\Delta_{k\sigma}+i\gamma)}.\label{eq:P2}
\end{equation}

For a medium of non-interacting molecules, because
all molecules contribute equally to the polarization, the polarization is obtained
from Eq.~(\ref{eq:P2}) by multiplying it with the molecular number
density $1/\upsilon_{0}$, i.e.
\begin{equation}
\vec{P}(t)=-\frac{1}{\hbar\upsilon_{0}}\mathrm{Re}\sum_{k,\sigma}\mbox{}^{\prime}\frac{\vec{d}_{g,k\sigma}\vec{d}_{k\sigma,g}\cdot\vec{E}(t)}{\omega-\Delta_{k\sigma}+i\gamma}.\label{eq:P3}
\end{equation}
For a double-ring M\"{o}bius molecule with $N$ carbon atoms in each ring,
the radius is $R\simeq NR_{c}/\pi$ and the width $4W=4R_{c}$, where
$R_{c}=0.077$nm \cite{Silbey04} is the radius of a single carbon atom. As a result,
the volume occupied by a single M\"{o}bius molecule is
\begin{eqnarray}
\upsilon_{0} & \simeq & 2\pi(R+W)^{2}W.
\end{eqnarray}

The magnetization is derived in analogy with Eq.~(\ref{eq:P3}) by
noticing the similarity between $H_{B}^{\prime}$ and $H_{E}^{\prime}$,
i.e.,
\begin{eqnarray}
\langle \vec{m}\rangle  & = & -\mathrm{Re}\sum_{k,\sigma}\mbox{}^{\prime}\frac{\vec{m}_{g,k\sigma}\vec{m}_{k\sigma,g}\cdot\vec{B}(t)}{\hbar(\omega-\Delta_{k\sigma}+i\gamma)},\label{LRFT magnetic dipole}\\
\vec{M}(t) & = & -\frac{1}{\hbar\upsilon_{0}}\mathrm{Re}\sum_{k,\sigma}\mbox{}^{\prime}\frac{\vec{m}_{g,k\sigma}\vec{m}_{k\sigma,g}\cdot\vec{B}(t)}{\omega-\Delta_{k\sigma}+i\gamma}\nonumber \\
 & = & -\frac{\mu_{0}}{\hbar\upsilon_{0}}\mathrm{Re}\sum_{k,\sigma}\mbox{}^{\prime}\frac{\vec{m}_{g,k\sigma}\vec{m}_{k\sigma,g}\cdot\vec{H}(t)}{\omega-\Delta_{k\sigma}+i\gamma}.
\end{eqnarray}
Here, $\vec{m}_{k\sigma,g}=\langle k,\sigma|\vec{m}|g\rangle $
is matrix element of the magnetic dipole operator.

\section{Local Field Correction}
\label{Appendix_Add}

Influenced by polarization of nearby molecules, the actual field that is exerted on an individual molecule could be different from the applied external field. Consequently, the total field at the molecule is modified as
\begin{equation}
\vec{E}_{\textrm{tot}}=\vec{E}+\vec{E}_i,
\end{equation}
where $\vec{E}$ and $\vec{E}_i$ denote respectively the external and internal fields, and the latter is due to the effect caused by the surrounding molecules. Although the internal field can be obtained by mean-field approximation as
\begin{equation}
\vec{E}_{\textrm{mean}}=-\frac{1}{V}\int_V d\vec{r} \nabla \Phi(\vec{r}),
\end{equation}
where $\Phi(\vec{r})$ is the scalar potential due to charge distribution inside a volume $V$, the approximation fails to take the effect of arrangement of the closest molecules into account. The volume is properly chosen to include a macroscopic number of molecules but still such small enough that the dipole moments for each molecules inside are approximately the same. To account for the arrangement of nearby molecules, the internal field is evaluated by the microscopic contribution $\vec{E}_{\textrm{near}}$ from all charges of the nearby molecules in $V$ subtracted by the mean-field part. This yields \cite{Jackson99}
\begin{equation}
\vec{E}_i=\vec{E}_{\textrm{near}}-\vec{E}_{\textrm{mean}}.
\end{equation}
However, for molecules arranged with higher symmetry, e.g., cubic lattice, or totally disordered, the contribution from $\vec{E}_{\textrm{near}}$ can be neglected \cite{Jackson99}. Furthermore, by expanding the scalar potential up to the dipole correction and assuming no net charge in the volume $V$, the mean-field part is evaluated as \cite{Jackson99}
\begin{equation}
\vec{E}_{\textrm{mean}}=-\frac{1}{3\varepsilon_0}\sum_{l}\frac{\vec{p}_l}{V},
\end{equation}
where $\vec{p}_l$ denotes the induced dipole moment of $l$th molecule inside the volume $V$.

For a sufficiently weak field, the induced dipole moment is proportional to the exerted field, i.e.
\begin{equation}
\vec{p}_l=\varepsilon_0 \gamma_{\textrm{mol}} \vec{E}_{\textrm{tot}},\label{E5}
\end{equation}
where the molecular polarizability $\gamma_{\textrm{mol}}$ is deduced according to Eq.~(\ref{eq:P2}) as
\begin{equation}
\gamma_{\textrm{mol}}^{(ij)}=-\frac{1}{\hbar \varepsilon_0}\sum_{k,\sigma}\mbox{}^{\prime}\frac{d_{g,k\sigma}^{(i)}d_{k\sigma,g}^{(j)}(\omega-\Delta_{k\sigma})}{(\omega-\Delta_{k\sigma})^2+\gamma^2}.
\end{equation}
As the volume $V$ is chosen such that its enclosed $\vec{p}_l$s are approximately the same, the polarization within the volume is related to the dipole moment of a representative molecule $\vec{p}_1$, multiplied with the molecule density $1/\upsilon_0$
\begin{eqnarray}
\vec{P}=\sum_l\frac{\vec{p}_l}{V}=\frac{1}{\upsilon_0}\vec{p}_1.\label{E7}
 \end{eqnarray}
On the other hand, the polarization $\vec{P}$ is related to the external field through the electric susceptibility $\vec{P}=\varepsilon_0 \chi_e \vec{E}$. By combining Eqs.~(\ref{E5},\ref{E7}), we have
\begin{equation}
\varepsilon_0 \chi_e \vec{E}=\frac{\varepsilon_0}{\upsilon_0}\gamma_{\textrm{mol}}(\vec{E}+\frac{1}{3\varepsilon_0}\varepsilon_0 \chi_e \vec{E}),
\end{equation}
from which the electric susceptibility could be obtained. Furthermore, by $\overleftrightarrow{\varepsilon_r}=1+\chi_e$, the relative permittivity with the local field correction is given by
\begin{eqnarray}
\overleftrightarrow{\varepsilon_r}&=&1+(1-\frac{1}{3\upsilon_0}\gamma_{\textrm{mol}})^{-1}\frac{1}{\upsilon_0}\gamma_{\textrm{mol}}\nonumber \\
&=&\left[\begin{array}{ccc}
\frac{3-2\eta'}{3+\eta'} & 0 & 0\\
0 & \frac{3+2\eta'}{3+5\eta'} & -\frac{6\eta'}{3+5\eta'}\\
0 & -\frac{6\eta'}{3+5\eta'} & \frac{3-7\eta'}{3+5\eta'}
\end{array}\right], \label{eq:16}
\end{eqnarray}
where in the second line an approximation similar to the one for Eq.~(\ref{ep}) has been used. It follows from Eq.~(\ref{eq:16}) that the corresponding eigenvalues of $\overleftrightarrow{\varepsilon_r}$ are given by $(3-2\eta^\prime)/(3+\eta^\prime)$, $(3-10\eta^\prime)/(3+5\eta^\prime)$ and $1$ respectively.

Compared with the results without the local field correction, the bandwidth does not change qualitatively. From relative permittivity, the bandwidth near the lowest inter-band transition frequency $\Delta_{0,\uparrow}$ without local field correction is limited by the separation between two zeros of $1-5\eta^\prime(\omega)=0$, while by including the local field correction, it is limited by two zeros of $1-10\eta^\prime(\omega)/3=0$. Hence the bandwidth in those two cases are only differed by a small factor and thus the bandwidth of negative permittivity is not significantly modified when the local field effect is taken into consideration. Similar proof can also be applied to analysis of the bandwidth of negative permeability. In conclusion, the local field effect would not substantially modify the bandwidth of negative refraction.

\section{Poynting Vector and the Wave Vector Surface}
\label{Appendix5}

In this section, we follow the treatment in Ref.~\cite{Landau95} to prove
that Poynting vector is parallel or anti-parallel to the normal of the wave vector surface.
Let $\vec{n}$ be a point on the wave vector surface
\begin{equation}
f(n_{x},n_{y},n_{z})=0,
\end{equation}
and $\delta\vec{n}$ be such a small change to $\vec{n}$ on the wave-vector
surface that $\delta\vec{n}$ can be viewed as a vector in the
tangent plane at that point, i.e., $\delta\vec{n}\cdot\mbox{ }\partial f/\partial\vec{n}=0$.
Here, $\partial f/\partial\vec{n}$ is the normal of the wave vector
surface at the point $\vec{n}$.

According to the Maxwell's equations, we have
\begin{eqnarray}
\vec{k}\times\vec{E}&=&\omega\mu_{0}\overleftrightarrow{\mu_{r}}\cdot\vec{H}, \\
\vec{k}\times\vec{H}&=&-\omega\varepsilon_{0}\overleftrightarrow{\varepsilon_{r}}\cdot\vec{E}.
\end{eqnarray}
In combination with $\vec{k}=\omega\sqrt{\varepsilon_{0}\mu_{0}}\vec{n}$, these yield
\begin{eqnarray}
\delta\vec{n}\times\vec{E}+\vec{n}\times\delta\vec{E}&=&c\delta\vec{B},\label{A2} \\
\delta\vec{n}\times\vec{H}+\vec{n}\times\delta\vec{H}&=&-c\delta\vec{D}.\label{A3}
\end{eqnarray}
By multiplying both sides of Eq.~(\ref{A2}) by $\vec{H}$, we have
\begin{eqnarray}
c\vec{H}\cdot\delta\vec{B} & = & \vec{H}\cdot(\delta\vec{n}\times\vec{E})+\vec{H}\cdot(\vec{n}\times\delta\vec{E})\nonumber \\
 & = & \vec{S}\cdot\delta\vec{n}-\delta\vec{E}\cdot(\vec{n}\times\vec{H})\nonumber \\
 & = & \vec{S}\cdot\delta\vec{n}+c\delta\vec{E}\cdot\vec{D}.\label{A4}
\end{eqnarray}
Similarly, multiplying both sides of Eq.~(\ref{A2}) by $\vec{E}$ yields
\begin{eqnarray}
-c\vec{E}\cdot\delta\vec{D} & = & \vec{E}\cdot(\delta\vec{n}\times\vec{H})+\vec{E}\cdot(\vec{n}\times\delta\vec{H})\nonumber \\
 & = & -\vec{S}\cdot\delta\vec{n}-\delta\vec{H}\cdot(\vec{n}\times\vec{E})\nonumber \\
 & = & -\vec{S}\cdot\delta\vec{n}-c\delta\vec{H}\cdot\vec{B}.\label{A5}
\end{eqnarray}

It follows from Eqs.~(\ref{A4},\ref{A5}) that
\begin{equation}
\vec{S}\cdot\delta\vec{n}=\frac{1}{2}c[\vec{H}\cdot\delta\vec{B}-\delta\vec{H}\cdot\vec{B}
+\vec{E}\cdot\delta\vec{D}-\delta\vec{E}\cdot\vec{D}].\label{eqn4Sn}
\end{equation}
Because $\overleftrightarrow{\varepsilon_{r}}$ and $\overleftrightarrow{\mu_{r}}$
are independent on $\vec{n}$ \cite{Landau95},
\begin{eqnarray}
\vec{H}\cdot\delta\vec{B} & = & \vec{H}\cdot\overleftrightarrow{\mu_{r}}\cdot\delta\vec{H}\nonumber \\
 & = & \sum_{ij}H_{i}\mu_{r}^{(ij)}\delta H_{j}.
\end{eqnarray}
For \textit{symmetric} $\overleftrightarrow{\mu_{r}}$ and $\overleftrightarrow{\varepsilon_{r}}$, i.e.
\begin{equation}
\mu_{r}^{(ij)}=\mu_{r}^{(ji)},\mbox{ }\varepsilon_{r}^{(ij)}=\varepsilon_{r}^{(ji)},
\end{equation}
we have
\begin{eqnarray}
\vec{H}\cdot\delta\vec{B} & = & \sum_{ij}H_{i}\mu_{r}^{(ji)}\delta H_{j}\nonumber \\
 & = & \delta\vec{H}\cdot\overleftrightarrow{\mu_{r}}\cdot\vec{H}\nonumber \\
 & = & \delta\vec{H}\cdot\vec{B}.\label{A9}
\end{eqnarray}
Similarly, we can also prove that
\begin{eqnarray}
\vec{E}\cdot\delta\vec{D} & = & \sum_{ij}E_{i}\varepsilon_{r}^{(ij)}\delta E_{j}\nonumber \\
 & = & \sum_{ij}E_{i}\varepsilon_{r}^{(ji)}\delta E_{j}\nonumber \\
 & = & \delta\vec{E}\cdot\vec{D}.\label{A11}
\end{eqnarray}
Inserting Eqs.~(\ref{A9},\ref{A11}) into Eq.~(\ref{eqn4Sn}) yields
\begin{equation}
\vec{S}\cdot\delta\vec{n}=0.\label{A13}
\end{equation}
Equation~(\ref{A13}) indicates that the Poynting vector $\vec{S}$
is along the normal of the wave vector surface at given point $\vec{n}$.


\begin{thebibliography}{10}
\bibitem{Veselago68}V. G. Veselago, Sov. Phys. Uspekhi. \textbf{10},
509 (1968).

\bibitem{Pendry04}J. B. Pendry, Science \textbf{306}, 1353 (2004).

\bibitem{Leonhardt06}U. Leonhardt, Science \textbf{312}, 1777 (2006).

\bibitem{Pendry06}J. B. Pendry, D. Schurig, and D. R. Smith, Science
\textbf{312}, 1780 (2006).

\bibitem{Pendry00}J. B. Pendry, Phys. Rev. Lett. \textbf{85}, 3966
(2000).

\bibitem{Shen16}Y. Shen and Q. Ai, Sci. Rep. \textbf{6}, 20336 (2016).

\bibitem{Pendry99}J. B. Pendry, A. J. Holden, D. J. Robbins, and
W. J. Stewart, IEEE Trans. Microw. Theory Tech. \textbf{47}, 2075
(1999).

\bibitem{Smith00}D. R. Smith, W. J. Padilla, D. C. Vier, S. C. Nemat-Nasser,
and S. Schultz, Phys. Rev. Lett. \textbf{84}, 4184 (2000).

\bibitem{Padilla06}W. J. Padilla, D. N. Basov, and D. R. Smith, Mater.
Today \textbf{9}, 28 (2006).

\bibitem{Chang10}C. W. Chang, M. Liu, S. Nam, S. Zhang, Y. Liu, G.
Bartal, and X. Zhang, Phys. Rev. Lett. \textbf{105}, 235501 (2010).

\bibitem{McPhedran11}R. C. McPhedran, I. V. Shadrivov, B. T. Kuhlmey,
and Y. S. Kivshar, NPG Asia Mater. \textbf{3}, 100 (2011).

\bibitem{Soukoulis 11}C. M. Soukoulis and M. Wegener, Nat. Photonics
\textbf{5}, 523 (2011).

\bibitem{Chen05}Y. F. Chen, P. Fischer, and F. W. Wise, Phys. Rev.
Lett. \textbf{95}, 067402 (2005).

\bibitem{Thommen06}Q. Thommen and P. Mandel, Phys. Rev. Lett. \textbf{96},
053601 (2006).

\bibitem{Shen14}Y. Shen, H. Y. Ko, Q. Ai, S. M. Peng, and B. Y. Jin,
J. Phys. Chem. C \textbf{118}, 3766 (2014).

\bibitem{Orth13}P. P. Orth, R. Hennig, C. H. Keitel, and J. Evers,
New J. Phys. \textbf{15}, 013027 (2013).

\bibitem{Brechtefeld06}F. Brechtefeld, N. Lindlein, G. Leuchs, and
U. Peschel, Ring-Shaped Molecules as Split Ring Resonators of a Molecular
Metamaterial, in Photonic Metamaterials: From Random to Periodic,
Technical Dgest (CD) (Optical Society of America, 2006), paper WD21.

\bibitem{Heilbronner64}E. Heilbronner, Tetrahedron Lett. \textbf{5},
1923 (1964).

\bibitem{Zhao09}N. Zhao, H. Dong, S. Yang, and C. P. Sun, Phys. Rev.
B \textbf{79}, 125440 (2009).

\bibitem{Ajami03}D. Ajami, O. Oeckler, A. Simon, and R. Herges, Nature
\textbf{426}, 819 (2003).

\bibitem{Yoneda14}T. Yoneda, Y. M. Sung, J. M. Lim, D. Kim, and A.
Osuka, Angew. Chem. \textbf{126}, 13385 (2014).

\bibitem{Balzani08}V. Balzani, A. Credi, and M. Venturi, \textit{Molecular
Devices and Machines}:\textit{ Concepts and Perspectives for the Nanoworld}
(VCH-Wiley, Weinheim, 2008).

\bibitem{Byers61}N. Byers and C. N. Yang, Phys. Rev. Lett. \textbf{7},
46 (1961).

\bibitem{Guo09}Z. L. Guo, Z. R. Gong, H. Dong, and C. P. Sun, Phys.
Rev. B \textbf{80}, 195310 (2009).

\bibitem{Walba93}D. M. Walba, T. C. Homan, R. M. Richards, and R.
C. Haltiwanger, New J. Chem. \textbf{17}, 661 (1993).

\bibitem{Salem72}L. Salem, \textit{The Molecular Orbital Theory of
Conjugated Systems} (Benjamin, Reading, MA, 1972).

\bibitem{Ceulemans98}A. Ceulemans, L. F. Chibotaru, and P. W. Fowler,
Phys. Rev. Lett. \textbf{80}, 1861 (1998).

\bibitem{Jackson99}J. D. Jackson, \textit{Classical Electrodynamics}
3rd Ed., (John Wiley, United States, 1999).

\bibitem{Marques08}R. Marqu\'{e}s, F. Mart\'{i}n, and M. Sorolla, \textit{Metamaterials
with Negative Parameters: Theory, Design and Microwave Applications}
(John Wiley, New Jersey, 2008).

\bibitem{Kubo85}R. Kubo, M. Toda, and N. Hashitsume, \textit{Statistical
Physics II Nonequlibirum Statistical Mechanics} (Springer-Verlag,
Berlin Heidelberg, 1985).

\bibitem{Greenwood72}H. H. Greenwood, \textit{Computing Methods in Quantum Organic
Chemistry} (Wiley-Interscience, Germany, 1972).

\bibitem{Silbey04}R. J. Silbey, R. A. Alberty, and M. G. Bawendi, \textit{Physical Chemistry, 4th Ed.}
(John Wiley\&Sons, Hoboken, 2004).

\bibitem{Landau95}L. D. Landau, E. M. Lifshitz, and L. P. Pitaevskii,
\textit{Electrodynamics of Continuous Media 2nd Ed.,} Chapter 11,
(Butterworth Heinmann, Oxford, 1995).

\bibitem{Shen05}N. H. Shen, Q. Wang, J. Chen, Y. X. Fan, J. P. Ding,
H. T. Wang, Y. Tian, and N. B. Ming, Phys. Rev. B \textbf{72},
153104 (2005).

\bibitem{Hu02}L. B. Hu and S. T. Chui, Phys. Rev. B \textbf{66},
085108 (2002).

\bibitem{Tao12}P. W. Tao, L. S. Hua, and Q. Z. Liang, Chin. Phys.
Lett. \textbf{29}, 034102 (2012).

\bibitem{Depine04}R. A. Depine and A. Lakhtakia, Microw. Opt. Techn.
Lett. \textbf{41}, 315 (2004).

\bibitem{McCall02}M. W. McCall, A. Lakhtakia, and W. S. Weiglhofer,
Eur. J. Phys. \textbf{23}, 353 (2002).

\bibitem{Tokuji09}S. Tokuji, J.-Y. Shin, K. S. Kim, J. M. Lim, K.
Youfu, S. Saito, D. Kim, and A. Osuka, J. Am. Chem. Soc. \textbf{131},
7240 (2009).

\bibitem{Yoon09}Z. S. Yoon, A. Osuka, and D. Kim, Nat. Chem. \textbf{1},
113 (2009).

\bibitem{Estrada12}E. Estrada and Y. Sim\'{o}-Manso, Chem. Phys.
Lett. \textbf{548}, 80 (2012).

\bibitem{Mukamel95}S. Mukamel, \textit{Principles
of Nonlinear Optical Spectroscopy} (Oxford University Press, New York,
1995).

\bibitem{Hawkea09}L. Hawkea, G. Kalosakasa, and C. Simserides, Mol. Phys. \textbf{107}, 1755 (2009).

\end{thebibliography}
\end{document}